\documentclass{aa}  
\usepackage{natbib}
\usepackage{graphicx}
\usepackage{txfonts,bm}
\usepackage[]{hyperref}
\usepackage{fontawesome}
\hypersetup{
   colorlinks,
   citecolor=magenta,
   filecolor=blue,
   linkcolor=magenta,
   urlcolor=blue
}
\usepackage[utf8]{inputenc}
\usepackage[T1]{fontenc}

\usepackage{float}
\usepackage{subcaption}
\usepackage{array}
 
\usepackage{comment}

\usepackage[linesnumbered,ruled,vlined]{algorithm2e}

\SetCommentSty{mycommfont}
\SetKwRepeat{Do}{do}{while}
\SetKwFor{Rtreefor}{rtree\char`_for}{do}{endfor}

\begin{document}

   \title{A highly accurate drag solver for multi-fluid dust and gas hydrodynamics on GPUs}

   \author{L. Sewanou\inst{1}\thanks{E-mail: leodasce.sewanou@ens-lyon.fr}
          \and
          G. Laibe\inst{1,2}
          \and
          B. Commer\c{c}on\inst{1}
          }

   \institute{ENS de Lyon, CRAL UMR5574, Universit\'e Claude Bernard Lyon 1, CNRS, Lyon, F-69007, France
         \and
             Institut Universitaire de France
             }
   
   \date{}

  \abstract
   {Exascale supercomputing unleashes the potential for simulations of astrophysical systems with unprecedented resolution. Taking full advantage of this computing power requires the development of new algorithms and numerical methods that are GPU friendly and scalable. In the context of multi-fluid dust-gas dynamics, we propose a highly accurate algorithm that is specifically designed for GPUs.
   }
   {We developed a multi-fluid gas-dust algorithm capable of computing friction terms on GPU architectures to machine precision, with the constraint for the drag-time step to remain a fraction of the global hydrodynamic time step for computational efficiency in practice.}
   {We present a scaling-and-squaring algorithm tailored to modern architectures for computing the exponential of the drag matrix, enabling high accuracy in friction calculations across relevant astrophysical regimes. The algorithm was validated through the \textsc{Dustybox}, \textsc{Dustywave}, and \textsc{Dustyshock} tests.}
   {The algorithm was implemented and tested in two multi-GPU codes with different architectures and GPU programming models: Dyablo, an adaptive mesh refinement code based on the Kokkos library, and Shamrock, a multi-method code based on Sycl. On current architectures, the friction computation remains acceptable for both codes (below the typical hydro time step) up to 16 species, enabling a further implementation of growth and fragmentation. This algorithm might be applied to other physical processes, such as radiative transfer or chemistry.}
   {}


   \maketitle

\section{Introduction}

The tiny dust grains originating from the interstellar medium play a crucial role in its evolution at all scales (e.g. \citealt{Draine2003,Draine2007}). They are dynamically coupled to the gas via a drag term and can thus exchange momentum and affect the gas dynamics. They act as charge reservoirs, allowing them to interact with the local magnetic field. They also interact with radiation, which affects the light that reaches Earth, and they play a key role in the local thermal balance by significantly contributing to opacity. Additionally, they regulate local chemistry and serve as fundamental building blocks for the formation of planetary embryos. Consequently, to achieve quantitative accuracy, numerical simulations of structure formation increasingly incorporate the most realistic models for dust processing (e.g. the discussions by \citealt{Bai2010,LP2012b,Lombart2021,Moseley2025}).

A major challenge arises from the fact that, locally, dust distributions can vary significantly, particularly in terms of size, which ranges from submicron scales to millimeter-sized particles, as seen in protoplanetary disks (e.g. \citealt{Bae2023,Lesur2023}). This variability is referred to as the polydisperse distribution of dust. This polydispersity adds complexity to the evolution of the system by generating dynamics that do not appear when all grains are of a single size \citep{Bai2010,Laibe2014c}. For instance, in a protoplanetary disk, larger grains tend to drift inward, which in turn induces a radial outward drift of the gas as a result of feedback effects \citep{Dipierro2018}. This outward gas motion then carries along part of the dust distribution. The consequences of this polydispersity can be critical by favouring or preventing the development of instabilities that are necessary to concentrate dust and form the first planetary embryos (e.g. \citealt{Krapp2019,Paardekooper2020,Krapp2020,Paardekooper2021,Wu2024,Canas2024,Paardekooper2025a,Paardekooper2025b,Matthijsse2025}). Last but not least, the dust grain size distribution evolves during the formation of a star and planets via coagulation and fragmentation processes. The latter are regulated by the coupling of the dust grains with the gas and the magnetic fields, for which the dust grain size is key \citep[e.g.,][]{Guillet2020,Lombart2021}.
In these scenarios, the behaviour of the gas-dust mixture is highly sensitive to the nature and distribution of the dust grains. Precise and rigorous numerical modelling is required to accurately capture these effects.

Numerically, addressing the polydispersity of strong constraints in hydrodynamic solvers presents several challenges (e.g. \citealt{Bai2010,Hutchison2018,BenitezLlambay2019,Mentiplay2020}). On one hand, increasing the number of fields required to evolve in order to simulate mixtures with different dust sizes imposes significant constraints on the computation time, memory access, and data management. In this regard, the solution lies in algorithmic and software advancements. In particular, the advent of exascale computing offers the possibility of overcoming some of these challenges, albeit at the cost of significantly higher energy and environmental demands (e.g. \citealt{Wibking2022,Grete2023,Lesur2023b,TDC2025}). On the other hand, intense friction regimes associated with small grains impose strict constraints on time-integration schemes. One way to circumvent this difficulty is to reformulate the gas-grain mixture equations in a limit where its evolution can be treated as an equivalent diffusive process \citep{Hutchison2018,Lebreuilly2019,Mentiplay2020}. This method is particularly effective when all grains are well coupled with the gas, but it becomes inadequate when part of the distribution is not (e.g. \citealt{Tricco2017,Commercon2023}). Another approach involves using unconditionally stable implicit numerical schemes \citep{Bai2010,Benitez_Llambay_2019,Krapp_Benitez_2020,Huang_2022,Krapp_2024}. These schemes typically require significantly more computational resources than explicit schemes, however, and they often lead to a trade-off in accuracy for the sake of practical feasibility. This is contrary to the goal of a high-precision numerical treatment that is necessary for accurately simulating complex physical processes. For these reasons, we aim to design a dust algorithm compatible with new multi-GPU architectures. By properly handling the friction matrix decomposition, we ensure maximum precision of the drag calculation without significantly impacting the overall computation time.

The manuscript is organised as follows. Section \ref{Sec:method} presents the equations and the most recent description for dust- and gas-mixture hydrodynamics and  introduces our novel exponential solver. In Sect. \ref{Sec:test} we test our method with respect to other drag solvers in different regimes using standard gas- and dust-dynamics test. We also assess and discuss its performance on GPUs. Section \ref{Sec:conclusion} concludes the paper.

\section{Numerical method\label{Sec:method}}

    \subsection{Equations of the evolution of a dust and gas mixture}
We considered an astrophysical dust-gas mixture and modelled $N$ pressure-less dust fluids. Each dust fluid $i$ was coupled to the gas via a drag term parametrised by $T_{\mathrm{s},i}$, the stopping time of the $i$th dust species. In most astrophysical conditions,  the mean free path of
the gas exceeds the size of the grain by far. In this case, the
grain stopping time of dust species $i$ is given by \cite{epstein:24}
\begin{equation}
 T_{\mathrm{s},i} =\sqrt{\frac{\pi \gamma}{8}} \frac{\rho_{\mathrm{grain},i}}{\rho} \frac{s_{\mathrm{grain},i}}{c_\mathrm{s}}, 
\end{equation}
where $\gamma$ is the adiabatic index of the gas, $c_\mathrm{s}$ is the sound speed of the gas, and $s_{\mathrm{grain},i}$ and $\rho_{\mathrm{grain},i}$  are the radius and intrinsic density of the grain. This intrinsic grain density is different from the dust density $\rho_\mathrm{d}$ (see below).

Conservation of momentum was ensured by the back-reaction term onto the gas momentum. The equations of evolution for the mixture are given by

    \begin{equation}
        \frac{\partial \rho_{\mathrm{g}}}{\partial t} + \nabla \cdot \left[\rho_{\mathrm{g}}\Vec{v}_{\mathrm{g}}\right] = 0 ,
        \label{eq:GenEq_1}
    \end{equation}
    
    \begin{equation}
        \frac{\partial(\rho_{\mathrm{g}}\Vec{v}_{\mathrm{g}})}{\partial t} + \nabla \cdot \left[\rho_{\mathrm{g}} \Vec{v}_{\mathrm{g}} \otimes \Vec{v}_{\mathrm{g}} + {P}_{\mathrm{g}} \mathrm{\bm{I}}\right] - \rho_{\mathrm{g}} \bm{f}_{\rm g} = \sum_{i=1}^{N} \rho_{\mathrm{d},i} \frac{\Vec{v}_{\mathrm{d},i} - \Vec{v}_{\mathrm{g}}}{T_{\mathrm{s},i}},
        \label{eq:GenEq_2}
    \end{equation}
    
    \begin{multline}
                  \frac{\partial E_{\mathrm{g}}}{\partial t} + \nabla \cdot \left[ (E_{\mathrm{g}} + P_{\mathrm{g}}) \Vec{v}_{\mathrm{g}}\right] -\rho_\mathrm{g} \Vec{f}_{\mathrm{g}} \cdot \Vec{v}_\mathrm{g} = \sum_{i=1}^{N} \rho_{\mathrm{d},i} \frac{ \Vec{v}_{\mathrm{d},i} -  \Vec{v}_\mathrm{g}}{T_{\mathrm{s},i}} \cdot \Vec{v}_\mathrm{g} \\ + w \sum_{i=1}^{N} \rho_{\mathrm{d},i} \frac{( \Vec{v}_{\mathrm{d},i} -  \Vec{v}_\mathrm{g})^2}{T_{\mathrm{s},i}},
                  \label{eq:GenEq_3}
    \end{multline}
    
    \begin{equation}
        \frac{\partial \rho_{\mathrm{d},i}}{\partial t} + \nabla \cdot \left[\rho_{\mathrm{d},i} \Vec{v}_{\mathrm{d},i}\right] = 0,
        \label{eq:GenEq_4}
    \end{equation}
    
    \begin{equation}
        \frac{\partial(\rho_{\mathrm{d},i} \Vec{v}_{\mathrm{d},i})}{\partial t} + \nabla \cdot \left[\rho_{\mathrm{d},i} \Vec{v}_{\mathrm{d},i} \otimes \Vec{v}_{\mathrm{d},i} \right] -  \rho_{\mathrm{d},i} \Vec{f}_{\mathrm{d},i} = \rho_{\mathrm{d},i} \frac{\Vec{v}_{\mathrm{g}} - \Vec{v}_{\mathrm{d},i}}{T_{\mathrm{s},i}} ,
        \label{eq:GenEq_5}
    \end{equation}

where $\rho_\mathrm{g}$, $\Vec{v}_{\mathrm{g}}$, $P_\mathrm{g}$, and $E_\mathrm{g}$ denote the gas density, the gas velocity and gas pressure, and the total gas energy $E_\mathrm{g}=\frac{1}{2} \rho_\mathrm{g}\Vec{v}^2_\mathrm{g}+e$, respectively, with $e=P_\mathrm{g}(\gamma-1)$ the gas internal energy. The densities and velocities of the dust fluid $i$ are denoted $\rho_{\mathrm{d,i}}$ and $\Vec{v}_{\mathrm{d},i}$, respectively. $\mathrm{\bm{I}}$ is the matrix identity. The external forces specific to each phase are denoted $\bm{f}_{\rm g}$ and $\bm{f}_{\mathrm{d},i}$. $w$ is a control variable that parametrises frictional heating: $w=0$  turns the frictional heating off, and $w=1$ means that the entire dissipation is deposited into the gas.  Eqs.~(\ref{eq:GenEq_1}) -- (\ref{eq:GenEq_5}) can be written in the following compact form: 
    \begin{equation}
        \frac{\partial \bm{U}}{\partial t} + \nabla \cdot F\left(\bm{U} \right)+ \Vec{f}_{\rm src,ext} = \Vec{f}_{\rm drag},
        \label{eq:CompEq}
    \end{equation}
    where 
    \begin{equation}
        \bm{U} \equiv  \left[\rho_{\mathrm{g}}, \rho_{\mathrm{g}}\Vec{v}_{\mathrm{g}}, E_\mathrm{g}, \rho_{\mathrm{d},i}, \rho_{\mathrm{d},i}\Vec{v}_{\mathrm{d},i} \right],
        \label{eq:CompVar1}
    \end{equation}

    \begin{equation}
        F\left(\bm{U} \right) \equiv \left[\rho_{\mathrm{g}}\Vec{v}_{\mathrm{g}}, \rho_\mathrm{g} \Vec{v}_\mathrm{g} \otimes \Vec{v}_\mathrm{g} + {P}_{\mathrm{g}} \bm {I},\left(E_{\mathrm{g}} + P_{\mathrm{g}} \right)\Vec{v}_{\mathrm{g}}, \rho_{\mathrm{d},i}\Vec{v}_{\mathrm{d},i}, \rho_{\mathrm{d},i} \Vec{v}_{\mathrm{d},i} \otimes \Vec{v}_{\mathrm{d},i}\right],
        \label{eq:CompVar2}
    \end{equation}

    \begin{equation}
        \Vec{f}_{\rm src,ext} \equiv \left[0,\rho_{\mathrm{g}}\Vec{f}_{\mathrm{g}}, \rho_{\mathrm{g}}\Vec{f}_{\mathrm{g}} \cdot \Vec{v}_{\mathrm{g}}, 0, \rho_{\mathrm{d},i}\Vec{f}_{\mathrm{d},i} \right],
        \label{eq:CompVar3}
    \end{equation}
and 
    \begin{multline}
        \Vec{f}_{\rm drag} \equiv \left[0, \sum_{i=1}^{N} \rho_{\mathrm{d},i} \frac{\Vec{v}_{\mathrm{d},i} - \Vec{v}_{\mathrm{g}}}{T_{\mathrm{s},i}},\sum_{i=1}^{N} \rho_{\mathrm{d},i} \frac{ \Vec{v}_{\mathrm{d},i} -  \Vec{v}_{\mathrm{g}}}{T_{\mathrm{d},i}} \cdot \Vec{v}_{\mathrm{g}} \right. \\
        \left. + w \sum_{i=1}^{N} \rho_{\mathrm{d},i} \frac{( \Vec{v}_{\mathrm{d},i} -  \Vec{v}_{\mathrm{g}})^2}{T_{\mathrm{s},i}}, 0 ,\rho_{\mathrm{d},i} \frac{\Vec{v}_{\mathrm{g}} - \Vec{v}_{\mathrm{d},i}}{T_{\mathrm{s},i}} \right].
        \label{eq:CompVar4}
    \end{multline}

 In Eqs. (\ref{eq:CompEq}) -- (\ref{eq:CompVar4}), $\bm{U}$ denotes the state vector of conservative variables, and $ F\left(\bm{U} \right)$ is a rank-$2$ tensor, the rows of which are the fluxes of each component of $U$. $ \Vec{f}_{\rm src,ext}$ and $ \Vec{f}_{\rm drag}$  denote external source terms and drag source terms.

    \subsection{Review of current dust-gas integrators}
    \label{sec:integrators}
    
Equation~(\ref{eq:CompEq}) is usually solved by splitting the time step to separately integrate the transport term, the source terms, and the drag terms. Gas and dust are treated in a similar way, except for the energy equation. Drag integration requires solving the following equation:

    \begin{equation}
        \frac{\partial \Tilde{\bm{U}}}{\partial t} = 
            \begin{bmatrix}
                    -\sum_{i=1}^{N} \alpha_i ( \epsilon_i \Vec{M}_{\mathrm{g}} - \Vec{M}_{\mathrm{d},i}) \\
                    \alpha_1 (\epsilon_1 \Vec{M}_{\mathrm{g}} - \Vec{M}_{\mathrm{d},1}) \\
                    \alpha_2 (\epsilon_2 \Vec{M}_{\mathrm{g}} - \Vec{M}_{\mathrm{d},2}) \\
                    \vdots \\
                     \alpha_{N} (\epsilon_{N} \Vec{M}_{\mathrm{g}} - \Vec{M}_{\mathrm{d},N}) \\
                \end{bmatrix}
                \equiv \Vec{\Tilde{f}}_{\rm drag},
            \label{eq:DragEq}
    \end{equation}
where
    \begin{equation}
        \Tilde{\bm{U}} \equiv \left[ \rho_{\mathrm{g}}\Vec{v}_{\mathrm{g}}, \rho_{\rm d,1}\Vec{v}_{\mathrm{d},1},..., \rho_{\mathrm{d},N}\Vec{v}_{{\mathrm{d}},N}\right]^{T} = \left[\Vec{M}_{\mathrm{g}}, \Vec{M}_{\mathrm{d},1},..., \Vec{M}_{\mathrm{d},N} \right]^{T},
        \label{eq:DragEqVar1}
    \end{equation}

    \begin{equation}
        \epsilon_i \equiv \frac{\rho_{\mathrm{d},i}}{\rho_{\mathrm{g}}},
        \label{eq:DragEqVar2}
    \end{equation}    
    and 
    \begin{equation}
        \alpha_i \equiv \frac{1}{T_{\mathrm{s},i}}.
        \label{eq:DragEqVar3}
    \end{equation}

Assuming linear drag terms, Eq.~(\ref{eq:DragEq}) can be set in the form of a system of first-order ordinary differential equations,
    \begin{equation}
        \frac{\partial \Tilde{\bm{U}}}{\partial t} = \Vec{\Omega} \Tilde{U},
        \label{eq:DragEqBis}
    \end{equation}
where $\Vec{\Omega}$ is the $(N+1)$ x $(N+1)$ sparse matrix, defined by 

    \begin{equation}
        \Vec{\Omega} \equiv
        \begin{bmatrix}
            -\sum_{i=1}^{N} \epsilon_i \alpha_i  & \alpha_1  & \alpha_2 & \dots & \alpha_{N} \\
            \epsilon_1 \alpha_1 & -\alpha_1 & 0 & \dots & 0 \\
            \epsilon_2 \alpha_2 & 0 & -\alpha_2 & \dots & 0 \\
            \vdots & \vdots &\vdots &\ddots & \vdots\\
            \epsilon_{N} \alpha_{N} & 0 & 0 & \dots & -\alpha_{N}
        \end{bmatrix}.
        \label{eq:jacobian}
    \end{equation}

    Several methods have been proposed to solve Eq.~(\ref{eq:DragEqBis}). \citet{Benitez_Llambay_2019} and \cite{Krapp_Benitez_2020} proposed a fully implicit first-order method that involves a similar sparse matrix that can be inverted analytically with the complexity of $\mathcal{O}\left( N \right)$. \citet{Huang_2022} proposed several second-order methods that are explicit or implicit, where the implicit schemes involve dense matrices that are computationally expensive to invert. The diagonally implicit Runge-Kutta (DIRK) method of \citet{Krapp_2024} combines the advantages of the first- and second-order fully implicit methods.

    \subsection{Exponential solver}
\subsubsection{Series methods for the matrix exponential}\label{sec:series}

    We chose to use the analytical solution of Eq.~(\ref{eq:DragEqBis}),
    \begin{equation}
        \Tilde{\bm{U}}(t) =  \mathrm{e}^{t \Vec{\Omega}}\Tilde{\bm{U}}(t_{0}).
        \label{eq:ExacSolEqu}
    \end{equation}    
    to build an implicit solver that preserves the accuracy of the scheme. The key difficulty was to maintain reasonable computational costs. Several algorithms have been proposed for the computation of the matrix exponential  \citep{Moler_Clever}. For a given matrix $\bm{A}$, the naive approach consists of performing a Taylor expansion in the truncation order $m$ according to
      \begin{equation}
        \mathrm{e}^{\bm{A}} = \sum_{i=0}^{\infty} \frac{\bm{A}^{i}}{i!} \simeq \sum_{i=0}^{m} \frac{\bm{A}^{i}}{i!} \equiv T_m \left( \bm{A}\right) ,
        \label{eq:ExpMatEqu}
    \end{equation}
which is usually not accurate enough because it entails truncation errors (both arithmetic and series truncation) and round-off errors \citep{Moler_Clever}. An alternative method consists of using 
the $\left(p,q\right)$ Padé approximant of $\mathrm{e}^{\bm{A}}$, defined by
    \begin{equation}
        \bm{R}_{pq}\left( \bm{A}\right) = N_{pq}\left( \bm{A}\right)(\bm{D}_{pq})^{-1},
        \label{eq:PadeEqu}
    \end{equation}
    where 
    \begin{equation}
        \bm{N}_{pq}\left( \bm{A}\right) = \sum_{j = 0}^{p} \frac{(p+q-j)!p!}{(p+q)!j!(p-j)!}\bm{A}^{j} ,
    \end{equation}
    and 
    \begin{equation}
        \bm{D}_{pq}\left( \bm{A}\right) = \sum_{j = 0}^{q} \frac{(p+q-j)! q!}{(p+q)!j!(q-j)!}(-\bm{A})^{j}.
        \label{eq:Dpq}
    \end{equation}
    
    This approximation holds when $\bm{D}_{pq}\left(\bm{A} \right)$ is not singular. This property is a priori not satisfied for all pairs $(p,q)$. The Taylor series of order $m$ is just a $\left(m,0\right)$ Padé approximant. This method provides a much better approximation for a given order than the Taylor series  \citep[see Table 1 in][]{Moler_Clever}. The Padé approximation is also limited by its own truncation and round-off errors, however. Both algorithms can be used when $\vert \vert A \vert \vert < 1 $ \citep{Sastre2015}.
    
        \subsubsection{Scaling and squaring method} \label{scal_squ_sub_sec}
        
     The most widely used method for computing the matrix exponential is the so-called scaling and squaring method \citep{Moly_Higham_2010}. The method relies on the following identity:
     \begin{equation}
         \mathrm{e}^{\bm{A}} = \left(\mathrm{e}^{\bm{A}/n} \right)^n,
         \label{eq:ExpEq1}
    \end{equation}
    where $n$ is a non-negative integer. It consists of computing an approximation for the exponential of the scaled matrix $\mathrm{e}^{\bm{A}/n}$, and then squaring the result $n$ times to have the approximation of $\mathrm{e}^{\bm{A}}$. The scaling step is performed with a Padé or Taylor approximation, which requires that $\vert \vert\bm{A}\vert \vert$ is sufficiently small (i.e. $\vert \vert\bm{A}\vert \vert < 1$) to control round-off errors. 

The scaling step is commonly performed using a Padé approximation to achieve double precision in the institute of electrical and electronics engineers (IEEE) 754 standard at lowest computational cost \citep{SKAFLESTAD2009783, Al-Mohy2009, Higham2009}. This requires a matrix inversion, as shown by Eq.~(\ref{eq:PadeEqu}). It is possible to achieve almost similar precision without paying the cost of this matrix inversion by the Taylor series, however \citep{RUIZ2016370, Sastre2009EFFICIENTST}. 
The scale factor $s$ is defined as 
\begin{equation}
2^{s} \equiv n .
\label{eq:def_s}
\end{equation}
 
 Importantly, \citet{Fasi2019} extended these methods for arbitrary precision, which is of prime interest for our drag algorithm because the aim is to achieve the required accuracy at minimum computational cost. 
 
 \citet{Sastre2015} proposed a modified version of the Taylor-based scaling and squaring algorithm.
 The first modification is relative to the choice of parameters $m_k^{*}$ (optimal Taylor polynomial order), $k^{*}$ (optimal matrix product required for the Taylor polynomial evaluation), and $s^{*}$ (optimal squaring parameter), such that the total number of matrix products $k^* + s^*$ is smaller than or equal to the number for the standard Taylor evaluation.
The second modification they proposed is to neglect high-order terms of the Taylor polynomials when for the given unit round-off $u$ the accuracy of the final result is not affected.

With these two modifications, they proposed four new Taylor-based scaling and squaring algorithms. Although two of the algorithms they proposed gave a comparable error to the expm routine of MATLAB (matrix exponential routine of MATLAB based on a Padé-approximation), however, we need to point out that all their algorithms are memory-bound. The reason is the need of storing a sequence of the matrix power $\{ A, A^2, \dots, A^q\}$, with $q$ a given integer known at runtime. The memory required to store the sequence of matrices for every cell in our simulation renders this method unfeasible on both GPUs and CPUs.
In \citet{Sastre2015}, the Taylor algorithm consisted of two parts:
    \begin{enumerate}
        \item A Paterson-Stockmeyer coefficient evaluation \citep[see ALGORITHM PS-COEFF in][]{Sastre2015}, and
        \item a Horner-Paterson-Stockmeyer evaluation of the order-$m$ Taylor polynomial \citep[see ALGORITHM HPS-EVAL in][]{Sastre2015}.
    \end{enumerate} 
We merged the two algorithms to reduce the memory footprint. The algorithmic details are presented in Appendix ~\ref{app:algo_exp}.

    \subsection{Implementation in Dyablo and Shamrock} \label{impl_dya_sham_sub_sec}
The different drag solvers we described above are all implemented in the codes Dyablo \citep{Dyablo} and Shamrock \citep{Shamrock}. These are two multi-GPU codes for astrophysical hydrodynamics simulations with the aim of targeting exascale supercomputers. They are both written in C++ (currently, C++17) and use MPI for the domain decomposition. For the shared memory parallelism, Dyablo uses Kokkos \citep{Kokkos3}, and Shamrock uses Sycl \citep{sycl2020}. Dyablo supports Eulerian AMR grid-based method. Shamrock supports both Lagrangian smoothed particle hydrodynamics (SPH) and Eulerian adaptive-mesh-refinement(AMR) grid-based methods. 

As pointed out in Sect. \ref{sec:integrators}, Eq.~(\ref{eq:CompEq}) --~(\ref{eq:CompVar4})
 was solved by the splitting technique.
 To evolve $\bm{U}^n $ from time $t=t^n$ to a new value $\bm{U} ^{n+1}$ at $t=t^{n+1}$,
 we first solved 
 \begin{equation}
     \frac{\partial \bm{U}}{\partial t} + \nabla \cdot F\left(\bm{U} \right)+ \Vec{f}_{\rm src,ext} =0,
     \label{eq:split_hyper}
 \end{equation}
 with $\bm{U}\left(t^n \right) = \bm{U}^n $ as initial condition, to obtain an intermediate state value $\Bar{\bm{U}}^{n+1}$. Then, using $\Bar{\bm{U}}^{n+1}$ as initial condition, we solved 
 \begin{equation}
        \frac{\partial \bm{U}}{\partial t} = \Vec{f}_{\rm drag}.
            \label{eq:split_drag}
\end{equation}

We integrated Eq.~(\ref{eq:split_hyper}) using a second-order Godunov method, whereas Eq.~(\ref{eq:split_drag}) was solved by one of the different drag solvers we described in Sect. \ref{sec:integrators}.

\section{Tests and performance\label{Sec:test}}
In this section, we present the numerical tests we performed to validate the method. We also present a performance analysis of our exponential drag solver within both Dyablo and Shamrock.

    \subsection{\textsc{Dustybox} test}
    \label{sec:collisions}
In this first test, we focused on the accuracy and the robustness of our matrix exponential solver alone (i.e. without the hydrodynamical solver step). The $1\rm{D}$ dust-gas \textsc{collisions} test \citep{Benitez_Llambay_2019} is a derivative of the \textsc{Dustybox} test \citep{LP11}, which consists of a system of gas and an arbitrary number of dust fluids with constant stopping times. The system evolves only under the momentum exchange between gas and dust fluids. In other words, we  integrated Eq.~(\ref{eq:DragEqBis}) for a given initial condition $ \Tilde{\bm{U}}_{0}$. \cite{Benitez_Llambay_2019} showed that the velocity of all fluids (gas and dust) is expected to converge to an asymptotic steady-state velocity (the velocity of the centre of mass), given by
\begin{equation}
\Vec{V}\left(t\right) = \Vec{V}_\mathrm{CM}+ \sum_{i=1}^{N}  \Vec{c}_{i} e^{\lambda_{i}t},
\label{eq:V_analytic}
\end{equation}
where $\Vec{V}_\mathrm{CM}$ is the velocity of the centre of mass, defined as
\begin{equation}
\Vec{V}_\mathrm{CM} = \frac{\Vec{v}_{\mathrm{g}} +  \epsilon_i \Vec{v}_{\mathrm{d},i} }{1+\sum_{i=1}^{N} \epsilon_i},
\label{eq:V_cm}
\end{equation}
and $\lambda_{i} < 0 $ is obtained by solving the eigenvalue problem,
\begin{equation}
\Vec{\Omega} \Vec{\bar{x}} = \Vec{\lambda} \Vec{\bar{x}}.
\label{eq:eigen_problem}
\end{equation}
The associate vector $\Vec{c}_{i}$ is then obtained by expressing the initial conditions into the eigenvectors.

We performed tests B (stiff case with short stopping times) and C (stiff case with high dust-to-gas ratios) described by \citet{Huang_2022} for our EXP \footnote{For the EXP integrator, we used the scaling and squaring method to compute $\mathrm{e}^{\Delta t \Vec{\Omega}}$ for each cell, where $\Omega$ is given by Eq.(\ref{eq:jacobian}).}  and different drag solvers of reference. In Table~\ref{tab:dust_collid_tab} we recall the initial conditions, the eigenvalues, and the coefficients for the analytical solution we used. 

\begin{table}[h]
\centering
 \begin{tabular}{c c}
      \begin{tabular}{c c c}
        \hline \hline
    Init.   &  B & C  \\
    \hline  
    $\Vec{\rho}_{g}$ & 1 & 1   \\
    $\Vec{v}_{\mathrm{g}}$ & 1 & 1 \\
    $\Vec{\rho}_{\mathrm{d,1}}$ & 1 & 10 \\
    $\Vec{v}_{\mathrm{d,1}}$ & 2 & 2 \\
    $T_{\mathrm{s},1}$ & 0.01 & 2 \\
    $\Vec{\rho}_{\mathrm{d,2}}$ & 1 & 100 \\
    $\Vec{v}_{\mathrm{d,2}}$ & 0.5 & 0.5 \\
    $T_{\mathrm{s},2}$ & 0.002 & 1 \\  
    \hline
    \end{tabular}
    &
        \begin{tabular}{c c c}
            \hline \hline 
        Coefs. & B & C \\
       \hline $V_{CM}$ & 1.16667 & 0.63964\\
        $\lambda_{1}$ & -141.74243 & -0.52370 \\
        $\lambda_{2}$ & -1058.25757 & -105.97630\\
        $c_{g,1}$ & -0.35611 & -0.06458\\
         $c_{g,2}$ & 0.18944 & 0.42494 \\
         $c_{d,1,1}$ & 0.85310 & 1.36237\\
         $c_{d,1,2}$ & -0.01977 & -0.00201\\
         $c_{d,2,1}$ & -0.49700 & -0.13559\\
         $c_{d,2,2}$ & -0.16967 & -0.00405\\  
    \hline
        \end{tabular}

 \end{tabular}
    \caption{Initial conditions (left) , eigenvalues, and coefficients for the analytical solution of the collision tests (right).}
    \label{tab:dust_collid_tab}
\end{table}

\begin{figure*}
\centering
\includegraphics[width=\textwidth]{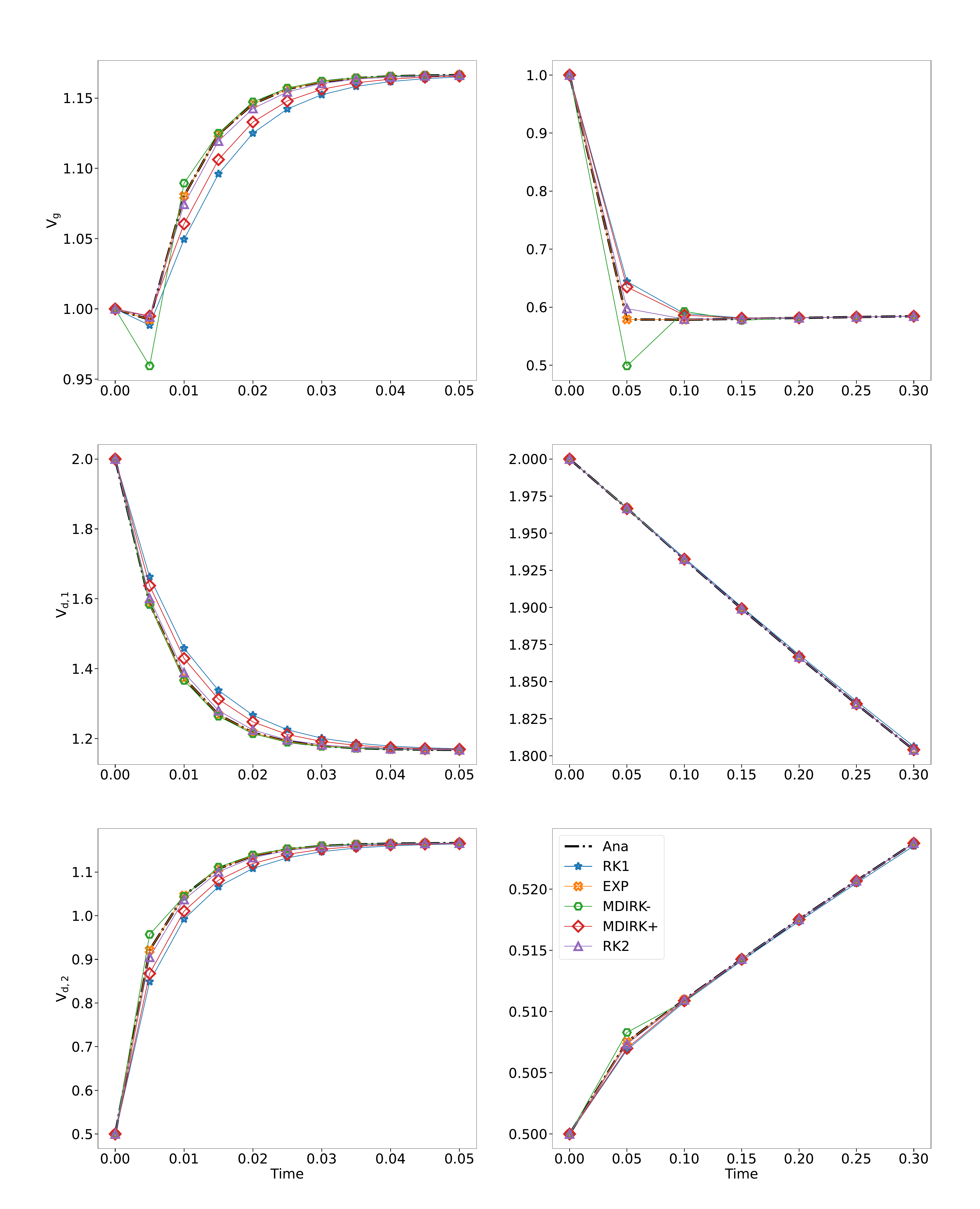}
\caption{\textsc{Collisions} test. Evolution of the gas and dust velocities (from top to bottom) for tests B and C (left and right columns, respectively). The dot-dashed black line represents the analytical solution. The other lines compare the numerical solutions obtained with the EXP integrator (orange) with other integrators, namely the first-order implicit RK1 (blue), the second-order fully implicit RK2 (purple), the MDIRK method with $\gamma = 1 - 1/\sqrt{2}$ (green), and the MDIRK method with $\gamma = 1 + 1/\sqrt{2}$ (red)}.
\label{fig:ColidTest_Fig}
\end{figure*}
Figure~\ref{fig:ColidTest_Fig} shows the evolution of velocities for \textsc{Collisions} tests B and C at a fixed time step of 0.005 and 0.05, respectively, for our EXP integrator (and others, presented in Sect. \ref{sec:integrators}). In both cases, the numerical velocity converges to the asymptotic steady-state velocity, and it matches the analytical solution well.

To quantify how well our matrix exponential solver performed in comparison with the drag solvers used in the literature, we performed a convergence analysis in time where the error was obtained following Eq.(40) of  \citet{Krapp_2024}, 
    \begin{equation}
        Er_1 \equiv \frac{1}{M} \sum_{k =0}^{2}\sum_{i=0}^{M-1} \frac{ |\tilde{v}_{k}(\Delta t_i)  -v_{k}(\Delta t_i)| }{v_{k}(\Delta t_i)},
        \label{eq:Error1_eq}
    \end{equation}
    with $v$ and $\tilde{v}$ denoting exact and numerical quantities, respectively. Here, $\Delta t_{i}$ is the time step, $M$ is the number of time steps during numerical integration, and $k=0,1,2$ is the index for the gas and dust fluids. Starting at $\Delta t = 10^{-4}$ ($10^{-3}$) for test B (for test C), we doubled the time step until we reached a final time of $t_\mathrm{final} = 10^{-1}$ ($2$) for test B and (test C). Here and in the remainder of the paper, we refer to RK1 as the first-order implicit drag solver, to RK2 as the second-order implicit Runge-Kutta solver, to MDIRK- and MDIRK+ as the versions of the multi-fluid second-order diagonally implicit Runge-Kutta solver, and to EXP as the exponential solver. 
    
\begin{figure*}
    \centering
    \includegraphics[width=\textwidth]
    {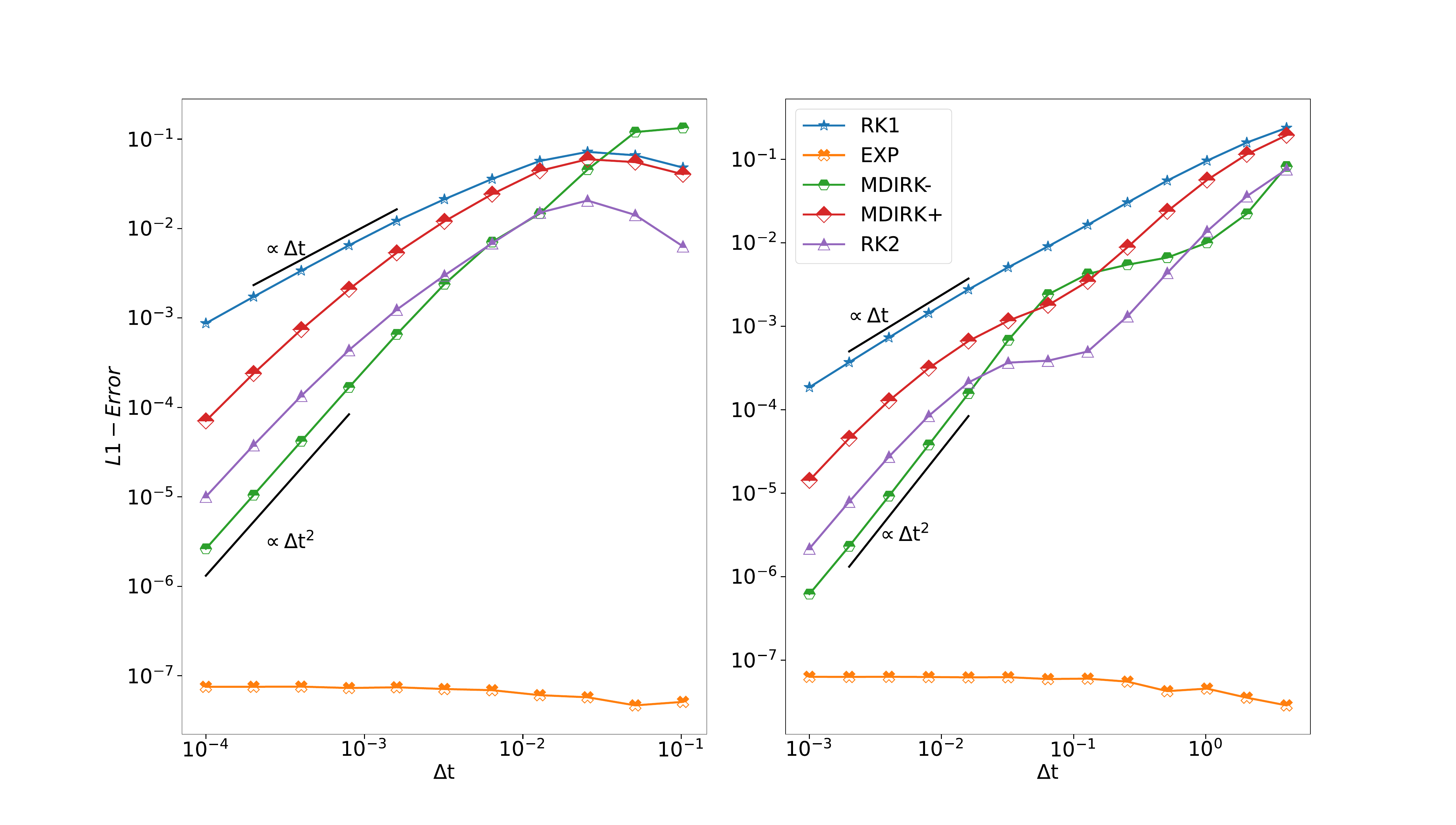}
    \caption{\textsc{Collisions} tests B (left) and C (right). The evolution of error $Er_1$ is defined by Eq.~(\ref{eq:Error1_eq}) for the EXP (orange) and other integrators: RK1 (blue), RK2 (purple), MKDIRK- (green), and MKDIRK+ (red).}
    \label{fig:colid_cvg_fig}
\end{figure*}

    Figure \ref{fig:colid_cvg_fig} shows the evolution of error $Er1$ as a function of the time step $\Delta t_i$. The RK1 and MDIRK+ solvers are first order, and the RK2 and MDIRK- solvers are second order. On the other hand, the error of the EXP matrix exponential solver remains almost constant around $10^{-8}$, regardless of $\Delta t$. This property arises because the choice of the optimal Taylor polynomial order $m_{k}^{*}$ and the optimal squaring was made such that the forward error was smaller than the unit roundoff $u$ (we set its value to $u = 2^{-53}$). This indicates that the EXP method is not affected by truncation errors and provides an accurate solution for the drag even at large time steps.

\begin{figure}
    \resizebox{\hsize}{!}{
    \includegraphics[width=\textwidth]{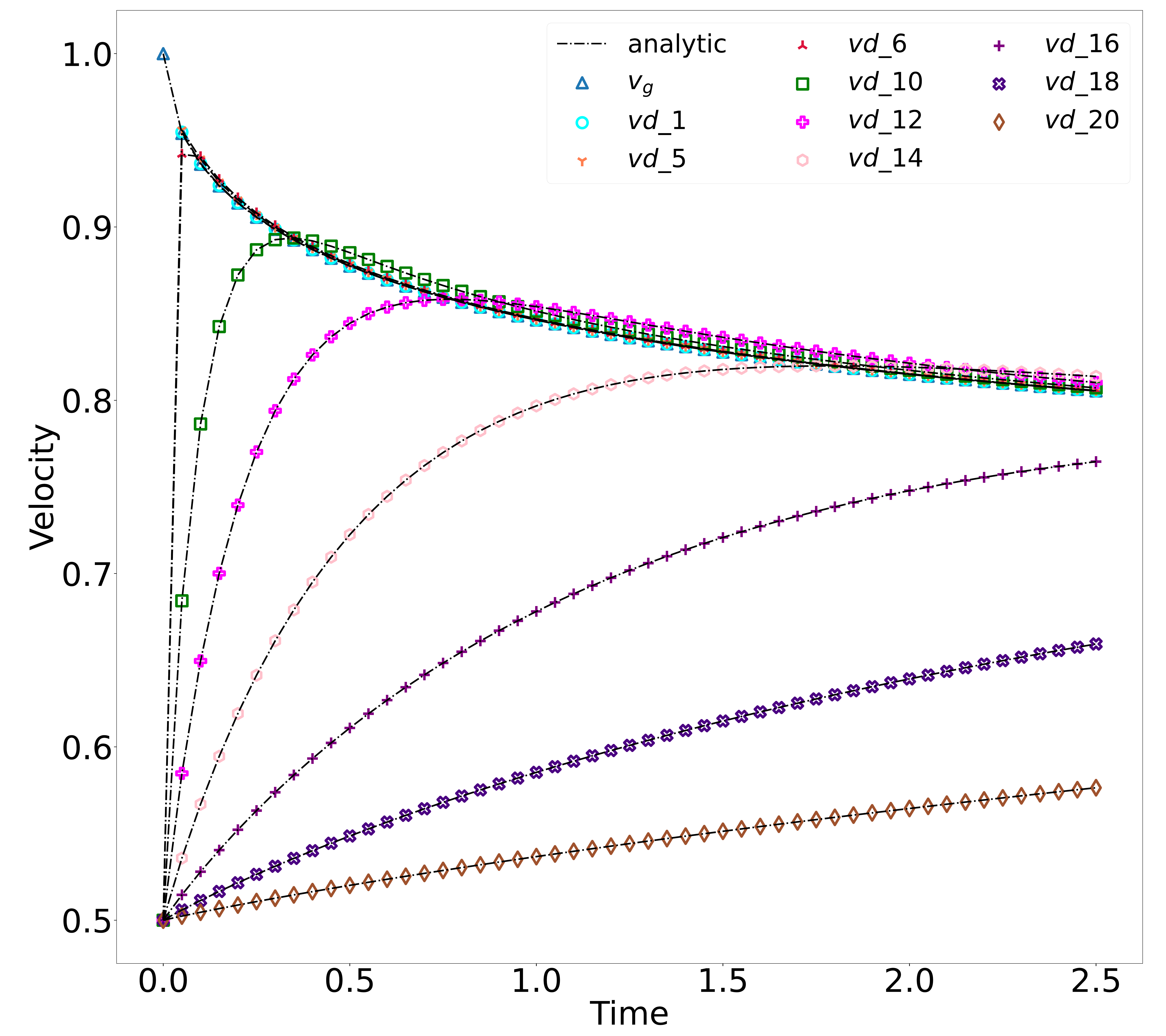}}
    \caption{Evolution of the gas and dust velocities for a \textsc{Collisions} test with 20 dust fluids obtained with the EXP solver and compared to the analytical solution given by Eq.~(\ref{eq:V_analytic}) (dashed black lines). For the first five dust species, their stopping time is short, such that they are strongly coupled to the gas. We therefore chose to plot the first and the fifth species. In addition to the $6$th dust species, we chose to skip every second fluid from the $10$th to the $20$th dust species.}
    \label{fig:collision_20_species_fig}
\end{figure}

To test the robustness of the matrix exponential solver, we also performed a collision test with $20$ dust fluids. The stopping time $T_{\mathrm{s},i}$ was uniformly spaced in log scale between $T_\mathrm{s,min} = 10^{-3}$ and $T_\mathrm{s,max} = 10$. The dust-to-gas ratios were given by 
\begin{equation}
    \epsilon_{i} =  \frac{\epsilon\left( \sqrt{T_{\mathrm{s},i+1}} - \sqrt{T_{\mathrm{s},i}}\right)}{\left( \sqrt{T_\mathrm{s,max}} - \sqrt{T_\mathrm{s,min}}\right)},
    \label{eq:dust_to_gas_ratio}
\end{equation}
such that $\epsilon = \sum_{i=1}^{N=20} \epsilon_{i} = 1.0$.
Figure~\ref{fig:collision_20_species_fig} shows the evolution of the gas and each of the 20 fluid velocities. Even in stiff regimes (short stopping times), the matrix exponential solver still performed well and provided errors of $\sim 10^{-7}$ for all dust fluids. 
Suppose for the RK1 scheme that $ Er_1 \propto \Delta t$ and the simulation time denotes $T_1  \propto \frac{1}{\Delta t}$, then to achieve an error of $\sim 10^{-7}$, the RK1 scheme requires $\sim 10^4$ ( $\sim 55$) more time on a \textsc{Collisions} test with two ( 20) dust fluids.

    \subsection{\textsc{Dustywave} test}
    \label{sec:dustywave}
The \textsc{Dustywave} test consisted in following the evolution of linear sound waves in a mixture of gas and dust. Because the dust phase is assumed to be pressure-less, it cannot support sound waves. The gas sound wave propagates with a modified sound speed in the mixture, however, because of gas pressure and momentum exchange. This test served as a benchmark  for the coupling between the transport and drag steps by solving Eqs.~(\ref{eq:GenEq_1}) -- (\ref{eq:GenEq_5}). The analytical solution for one gas and one dust fluid was proposed by \citet{Laibe_Price_2012} and was extended to an arbitrary number of dust fluids by \citet{Laibe2014c}.

We reproduced the $1\rm{D}$ test for one gas fluid with one and four dust fluids. The setup consisted of an isothermal sound wave ($P_\mathrm{g} = c_\mathrm{s}^2 \rho_\mathrm{g}$, with $c_\mathrm{s} = 1$) in a periodic box of size $L = 1$ and zero background velocity. The perturbation equation writes
\begin{equation}
    \delta f = A\left[ Re\left( \delta \hat{f}\right)\cos{\left(kx \right)} - Im\left( \delta \hat{f}\right)\sin{\left(kx \right)}\right],
    \label{eq:pert_1}
\end{equation}
with $A$ a small amplitude to ensure linearity ($10^{-4}c_\mathrm{s}$ and $10^{-4} \rho_\mathrm{g}^0$ for velocity and density perturbations), and with the wavenumber  $k= \frac{2\pi}{L}$. The test was performed on a uniform grid of $N_{\rm{cells}} = 128^{3}$. 

The initial background densities, perturbation amplitudes, and stopping were taken from Table 2 of \cite{Benitez_Llambay_2019}. 

\begin{figure*}
\centering
\includegraphics[width=\textwidth]{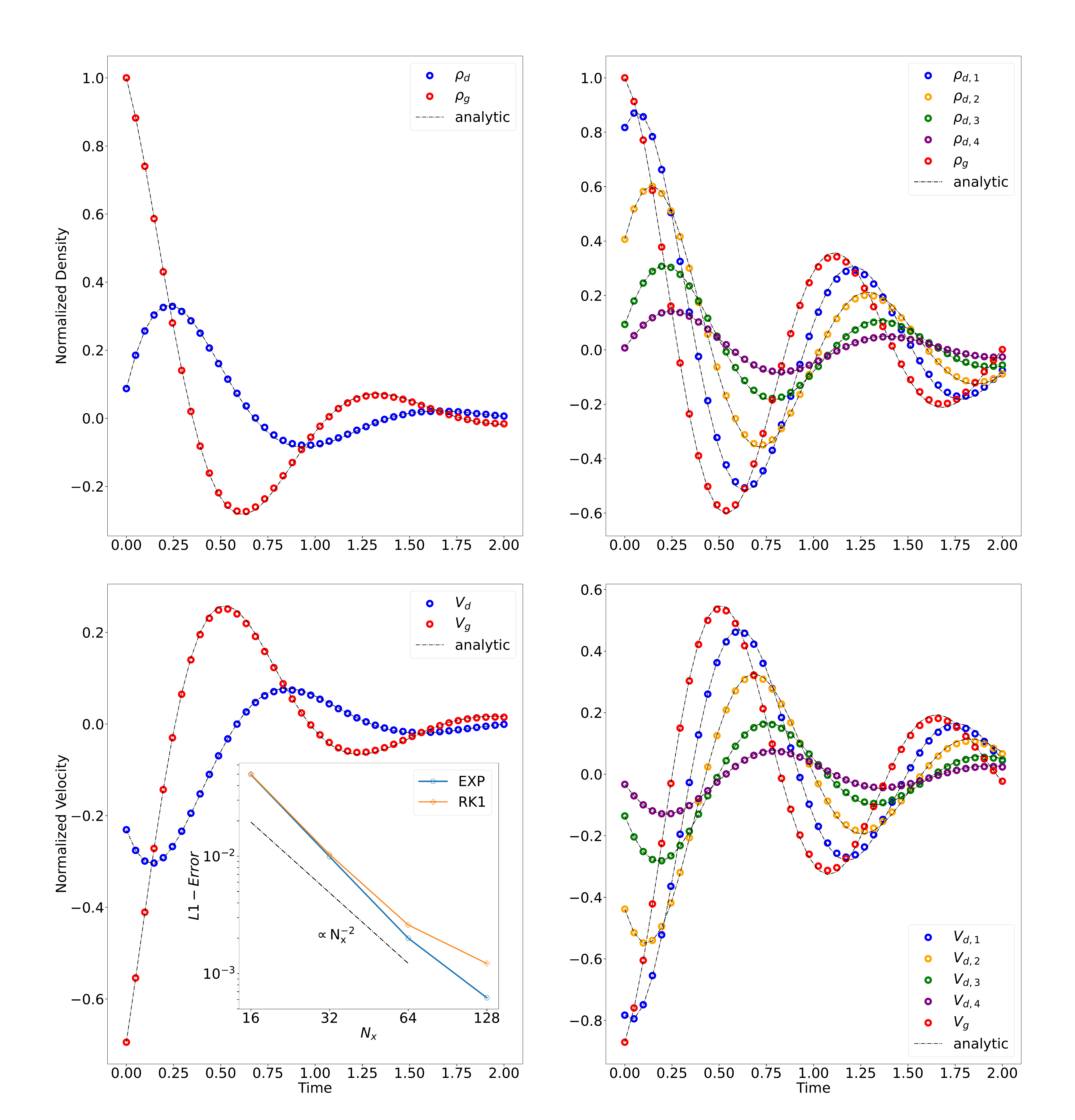}
\caption{\textsc{Dustywave} test. Evolution of normalised densities (upper panels) and velocities (lower panels) for one (left panel) and four dust fluids (right panel). The dashed black lines correspond to the analytical solution. The other markers and colours represent the numerical results obtained with the EXP algorithm. We additionally plot the L1 error given by Eq.~(\ref{eq:Error2_eq}) for the one-dust fluid case as an inset in the bottom left panel using a fixed time step of $\Delta t = 10^{-4}$ for EXP and RK1.}
\label{fig:dwave_Fig}
\end{figure*}

Figure~\ref{fig:dwave_Fig} shows the evolution of the normalised velocities $\left( \frac{\delta {\hat{v}}}{c_\mathrm{s} A}\right)$ and densities $\left(\frac{\delta {\hat{\rho}}}{\rho_0 A} \right)$ for two and five dust fluids, respectively. The numerical solutions recover the analytical solutions well (a few $0.1 - 1 \%$),  which indicates a  good coupling of the drag solver and the transport step. We additionally studied the convergence in space of the scheme (transport and drag together) by comparing the results with the RK1 solver, using a small time step $\Delta t = 10^{-4}$ such that the influence of the time error was minimised. For the transport step, we used a second-order MUSCL-Hancock scheme with the Minmod slope limiter. To compute the error, we used
Eq.~(46) from \citet{Krapp_2024} as follows:
\begin{equation}
    Er_2 \equiv \frac{1}{N_{\rm{cells}}} \sum_{k =1}^{2}\sum_{i=1}^{2} |\delta f_{i,\mathrm{numeric},k}  -\delta f_{i,\mathrm{analytic},k}| ,
    \label{eq:Error2_eq}
\end{equation}
with $k$ and $i$ the indices of the fluid and the fields (velocity and density), respectively.  The lower left panel of Fig.~\ref{fig:dwave_Fig} shows that the scheme has second-order accuracy in space, which is consistent with the order of the transport scheme we used. As the resolution increases, the EXP solver gives a better accuracy than the RK1 solver by a moderate factor.

    \subsection{\textsc{Dustyshock} test} \label{sec:dustyshock}
The dust-gas mixture shock test is a canonical benchmarking problem to test the robustness and accuracy of numerical codes for dusty gas. The steady-state J-type shock solution (when the shock speed exceeds the signal speed) in the case of a one-dust fluid, proposed by \cite{Lehm_Ward_2018}, was extended to an arbitrary number of dust fluids by \cite{Benitez_Llambay_2019}. 
The left state is given by
\begin{align}
     \rho_{\mathrm{g}}^- &= \rho_{\mathrm{g}0}, \nonumber \\
        \rho_{\mathrm{d},i}^- &= \epsilon_{i}\rho_{\mathrm{g}0}, \nonumber\\
        w_\mathrm{g}^-&= w_{\mathrm{d},i}= 1.0,
        \label{eq:eq_dustyshock_1}
\end{align}
and the asymptotic right states are given by
\begin{align}
    \rho_{\mathrm{g}/\mathrm{d},i}^+ &= \frac{\rho_{\mathrm{g}/\mathrm{d},i}^- w_{\mathrm{g}/\mathrm{d},i}^-}{w_{\mathrm{g}/\mathrm{d},i}^+}\rho_{\mathrm{g}0}, \nonumber\\
    w_{\mathrm{g}}^+ &= w_{\mathrm{d},i}^+ = \left( 1 + \sum_{i=1}^{N} \epsilon_{i}\right)^{-1} \mathcal{M}^{-2},
     \label{eq:eq_dustyshock_2}
\end{align}
where $w_{\mathrm{g}/\mathrm{d},i} = v_{\mathrm{g}/\mathrm{d},i}/v_\mathrm{s} $, $v_\mathrm{s}$ is the shock velocity, and $\mathcal{M} = v_\mathrm{s}/c_\mathrm{s}$ is the Mach number.
Following the same setup as in Sect. 3.3.2 from \cite{Benitez_Llambay_2019}, we performed the two-fluid and five-fluid tests. We list the initial conditions for the two-fluid and four-fluid tests below,
\begin{align}
        \left(K, \rho^-, \rho+, w^-, w^+ \right) &= \left(1.0,1.0,8.0,1.0,0.125 \right), \nonumber \\
    \left(K_1,K_2,K_3,\rho^-,\rho^+ ,w^-,w^+\right) &= \left( 1.0,3.0,5.0,1.0,16.0,1.0,0.0625\right), \nonumber \\
    \left(c_{\mathrm{s}},v_{\mathrm{s}},\epsilon, x_0,x_{\mathrm{start}},x_{\mathrm{end}},N_{\mathrm{cells}} \right) &= \left(1,2,1,4,0,40,512\right).
    \label{eq:dshock_init_cond_val}
\end{align}
$x_0$ is the initial jump position, and we used the zero-gradient boundary condition.
We used an HLLE  Riemann solver for the gas and the upwind solver of \cite{Huang_2022} for the dust. We used the van Leer slope limiter.

\begin{figure}
\resizebox{\hsize}{!}{
\includegraphics[width=\textwidth]{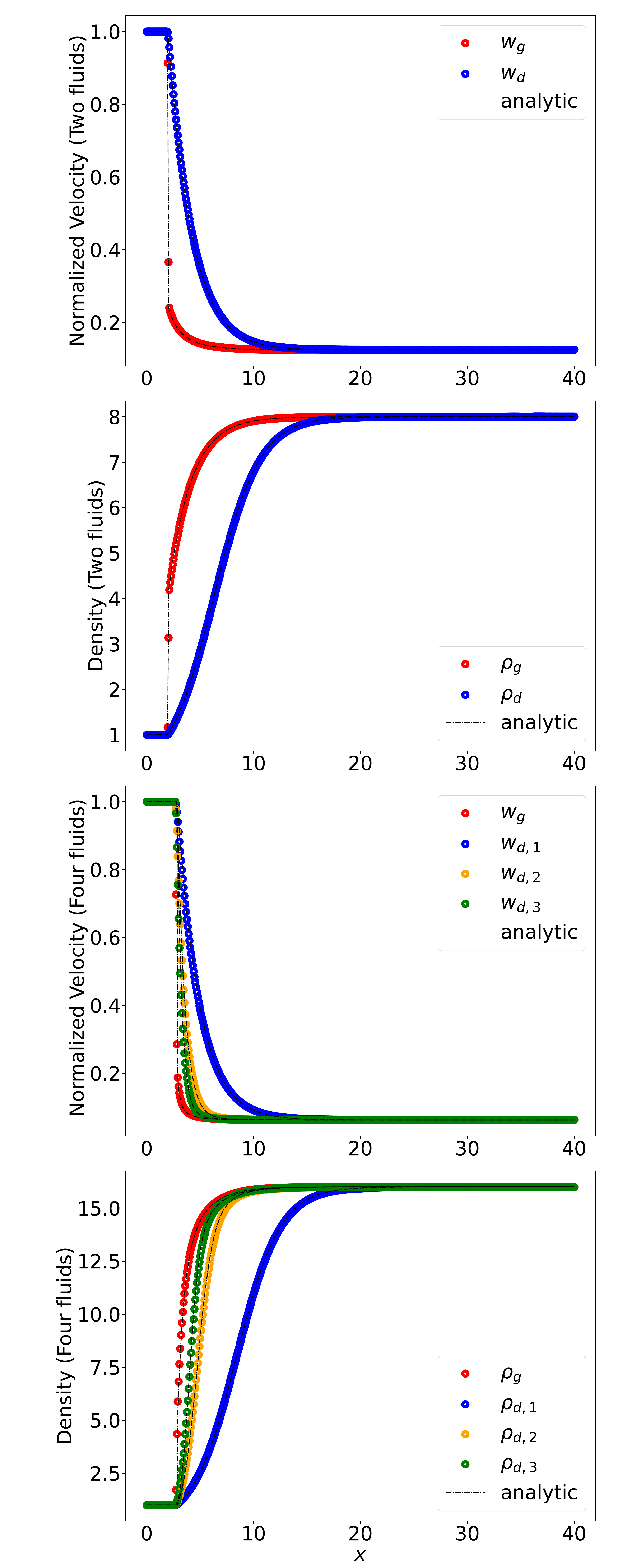}}
\caption{\textsc{Dustyshock} test. Normalised velocities (first and third panels for one- and three-dust fluid(s), respectively) and density (second and fourth panels for one- and three-dust fluid(s), respectively). The dashed lines show the analytical solutions, and the circle markers show the numerical results we obtained with the EXP algorithm. With the EXP solver, the shock is captured within one to two cells.} 
\label{fig:dshock_Fig}
\end{figure}

Figure~\ref{fig:dshock_Fig} shows the normalised velocities and densities  for gas and for one- (first two panels) and three- (last two panels) dust fluids at $t=500$. 

The good agreement between the analytical and numerical solutions ($\sim 0.1 - 1\%$) shows that our drag solver is correctly coupled to the transport step in a challenging regime.

    \subsection{Performance} \label{sec:perf}
We report the performance of our EXP algorithm on GPUs by performing the \textsc{Dustywave} test with various numbers of dust fluids. Following the same procedure as in Sect.~(\ref{sec:dustywave}), we generated uniform stopping times, computed the dust-to-gas ratio with Eq.~(\ref{eq:dust_to_gas_ratio}), solved the dispersion relation, and computed the associated eigenvectors \citep[see Eq. 45 -- 48 in][] {Benitez_Llambay_2019}.  The tests were performed on a NVIDIA DGX station with four A100 GPUs and an AMD 7742 CPU (64 cores, 2.25 - 3.4 GHz). By default, we used $N_{\mathrm{cells}} = 64^3$ and integrated until $t=2$. Profiling consisted in measuring the execution time of the relevant kernels using the tools Nsight compute and Nsight system of the NVIDIA Nsight developer tools suit. Although absolute performance is expected to improve with future hardware advancements and potential optimisation strategies, these tests provide useful estimates of the current tractability of the EXP algorithm to astrophysical simulations.

In the Dyablo code, the matrices required for the Taylor evaluation step are handled with the Kokkos array. These are static arrays, which implies that the code must be recompiled whenever the number of dust fluids changes. Because each cell needs to solve the drag equation, the cells of the overall grid are divided into blocks.  Within each block, a parallel\_for execution policy is applied in conjunction with a RangePolicy with a length equal to the block size, which allows each cell to integrate the drag equation. In the Shamrock code, the matrices are handled using mdspan, which are dynamic arrays that allow adjusting the number of dust fluids at runtime without recompilation. Within Shamrock, we experimented with two implementation  strategies. In the first implementation, mdspan arrays were attached to the device or global memory, and we used a basic data-parallel kernel. In the second implementation, we used ND-Range data-parallel kernels, where the number of work-items\footnote{In sycl, a work-item is the  smallest unit of execution on a GPU compute unit. It is the equivalent of the CUDA thread.}  per work-group\footnote{In sycl, a work-group is a set of work-items in the thread hierarchy that are scheduled together. It is the equivalent of warp with the CUDA programming model or wavefront for AMD GPU. The former's size is user-defined, while a warp or wavefront has a fixed size depending on the GPU architecture.} is computed with respect to the size of the shared memory available on the device and the number of dust fluids. In this case, mdspan arrays were directly attached to the local memory of the work-group.

\begin{figure*}
\centering
\includegraphics[width=\textwidth]{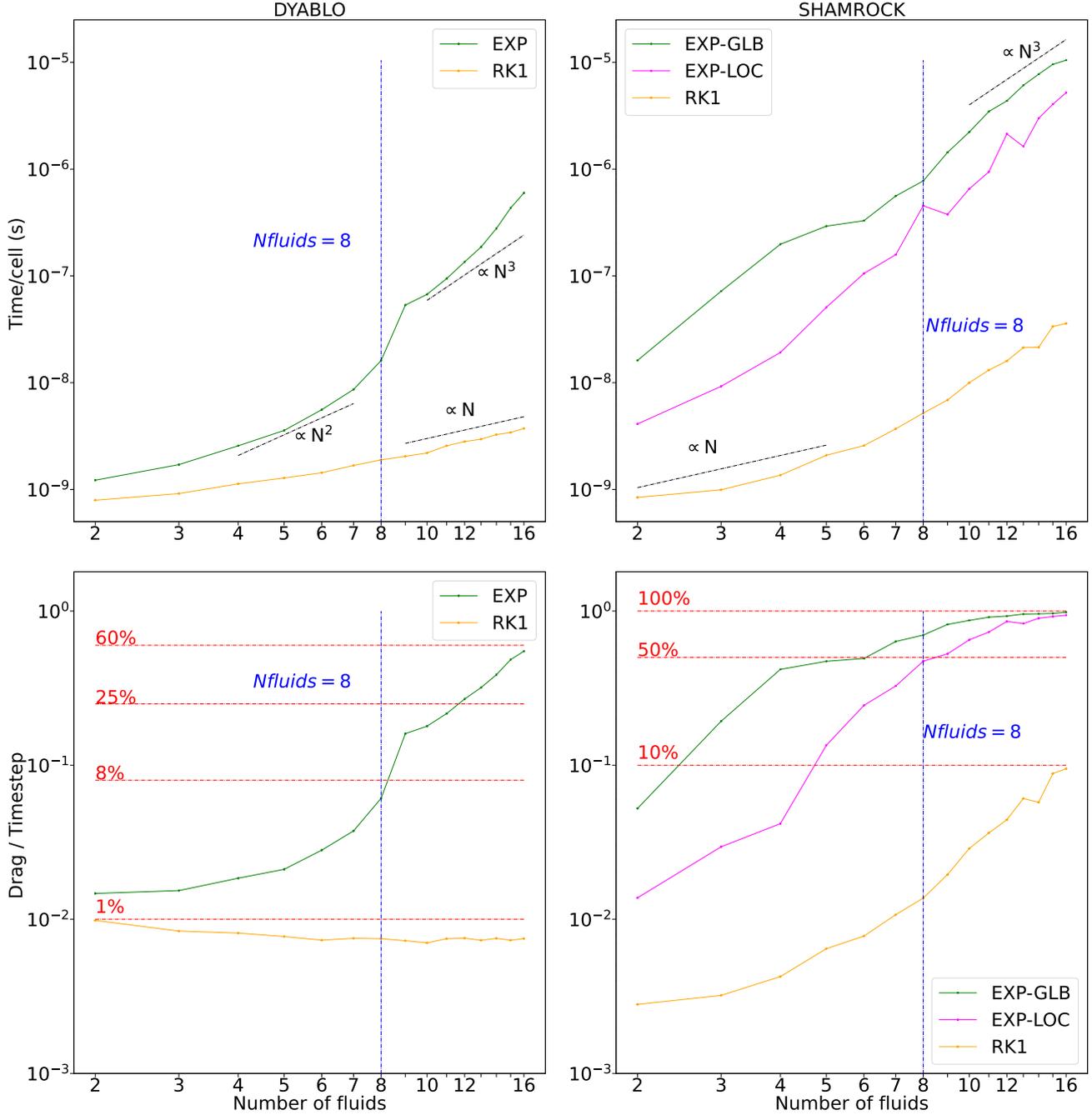}
\caption{Performance of of the EXP drag solver as function of the number of fluids (gas and dust fluids) on one A100 GPU obtained with an implementation with a moderate level of optimisation in the Dyablo (left) and the Shamrock (right) codes. We provide the results obtained with the RK1 algorithm for comparison. The tests with Dyablo used static arrays and global memory. The tests with Shamrock used dynamical arrays and either the local (EXP-LOC) or the global (EXP-GLB) memory. For 8-16 dust fluids, the computational cost of the drag step is $\sim 10 - 100\%$ of the hydro step, which makes it usable for astrophysical applications.
} 
\label{fig:perf_Fig}
\end{figure*}

Figure~\ref{fig:perf_Fig} shows the time per cell (top panels) and the relative cost of the drag kernels per time step (bottom panels) for both codes as function of the number of dust fluids $N$ for a resolution of $N_\mathrm{cells} = 64^3$ .  For reference, the computational cost of the RK1 solver scales as $\mathcal{O}\left( N \right)$, as expected, regardless of the implementation. For the EXP solver, the computational cost within Dyablo initially scales as $\mathcal{O}\left( N^2 \right)$ up to $8$ fluids and then becomes dominated by matrix-matrix operations. This results in a time complexity of $\mathcal{O}\left( N^3 \right)$. Within Shamrock, the local memory implementation of the EXP solver offers a slight performance improvement over the global memory implementation for small numbers of fluids (e.g. a factor of 10 for N = 4 ). As the number of fluids and thus the matrix sizes increases, the size of the work-groups decreases. Hence, the local and global memory implementations both have nearly identical execution times. One possible path for improvement might be to use hierarchical data-parallel kernel in which a work-group will compute the matrix exponential instead of assigning this task to a single thread, as is currently done. Furthermore, within Dyablo, the EXP solver takes $55\%$ of the overall time steps for up to 16 fluids, while with Shamrock, the drag time step with the same number of fluids is about the same as the hydro time step in both implementations. For instance, for up to 8 fluids, the EXP solver in Dyablo takes less than $8\%$, whereas the local memory version in Shamrock is slightly lower than $50\%$.

\begin{figure*}
\centering
\includegraphics[width=\textwidth]{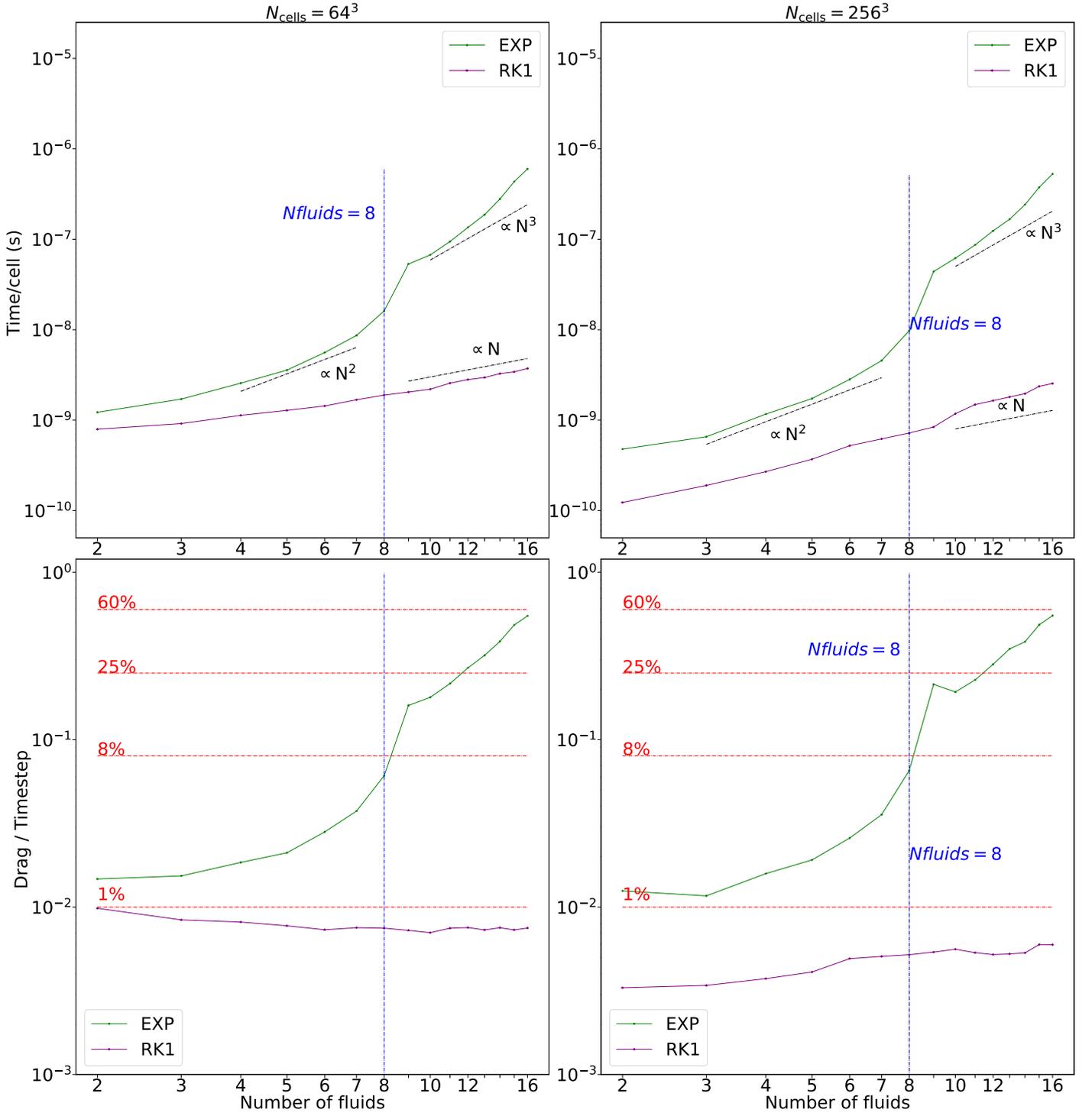}
\caption{Performances of the EXP drag solver on a \textsc{Dustywave} test performed on a A100 GPU using Dyablo for $N_\mathrm{cells} = 64^3$ (left) and $N_\mathrm{cells} = 256^3$ cells for 1 to 15 dust fluids, respectively. The top panels show the time per cell for the drag step. The bottom panels show the relative cost of the drag step with respect to the total time step (hydro+drag) per time step.} 
\label{fig:perf_Fig_dya}
\end{figure*}

Additionally, we analysed the effect of the resolution on the measured performances. Figure~\ref{fig:perf_Fig_dya} shows the time per cell (top panels) and the relative cost of the drag solver per time step (bottom panels) as a function of the number of dust fluids for $N_{\rm{cells}} =256^3$ (left panels) and $N_{\rm{cells}} = 64^3$ (right panels) cells (Dyablo code). We observe the same performance profile as for $N_{\rm{cells}} = 64^3$, where for EXP, we have a time complexity of $\mathcal{O}\left( N^2 \right)$ and $\mathcal{O}\left( N^3 \right)$ for $N \leq  8$ and $N \geq 9$, respectively. The computational cost with $N = 8 \left( \simeq 9.7\times10^{-9}~ \rm{s}\right)$  multiplied by about $4.5$ when moving to $N = 9 \left( \simeq 4.4\times10^{-8}~ \rm{s}\right)$ clearly shows the change in order in the time complexity. With the Nsight Compute profiler, both kernels have a large number of registers (the fastest memory storage area with a streaming multiprocessor) per thread, 256 each. There are  0.18 and 0.09  eligible warps (groups of threads  that are scheduled together and are executed in parallel) per cycle for $N=8$ and $N=9$ fluids, respectively, on average, however. For each warp, the latency (the amount of time wasted doing nothing) is 12.20 and 25.02 cycles on average between two consecutive executed instructions for 8 and 9 fluids, respectively. This shows that the actual implementation of the exponential solver is over using registers per thread, and as a result, it causes a low occupancy (a measure of the efficiency of the GPU computing resources). We provide the performances for the RK1 solver, which relies on an analytical prescription, for reference.

The above analyses suggest optimisation paths to make the EXP algorithm more competitive. A reduced pressure on the registers might lead to a better occupancy. Using several threads (e.g. a thread block) by matrix operation instead of a single thread, as is done currently, is worth considering. This might be done, for example, by using hierarchical parallelism. Finally, future GPUs with larger bandwidth and shared memory capacity (notably for the implementation in local memory) are also expected to reduce the latency of the algorithm.

\section{Conclusion\label{Sec:conclusion}}

We have presented and studied the EXP algorithm that we designed to compute the momentum exchange for dust-gas polydisperse mixtures very accurately on GPU architectures. The algorithm is based on a Taylor-based scaling and squaring matrix exponential computation. It was implemented in two different codes, namely Dyablo and Shamrock, using standard moderate optimisation strategies, and it was benchmarked on standard drag-dominated and hydro-dominated tests. The performance measurements showed that the cost of the drag time step does not significantly affect the code performance for up to $\sim$ 8-16 dust fluids.

The method is therefore suitable for astrophysical simulations even with growth and fragmentation as long as high-order solvers are used \citep{LL2021,Lombart2022,Lombart2024}. The algorithm can also be applied for linear or linearised source terms, such as in radiative transfer and chemistry.

\section*{Data availability}

The Dyablo code and the relevant sources are publicly distributed on Github \faGithub\ (\url{https://github.com/Dyablo-HPC/Dyablo}). The Shamrock code and the relevant sources are publicly distributed on Github \faGithub\,under the open source CeCILL v2.1 license (\url{https://github.com/Shamrock-code/Shamrock}).

\begin{acknowledgements}
We thank T. David-$\,\!$-Cléris, A. Durocher, M. Delorme, G. Lesur, U. Lebreuilly, G. Verrier, P. Hennebelle, F. Lovascio for useful comments and discussions, P. Huang for helping with the RK2 test, E. Quemener and the Centre Blaise Pascal de Simulation et de Modélisation Numérique for support on the benchmarks, the ENS de Lyon for partly funding the DGX Nvidia Workstation and the electric costs associated with local simulations.
 We ackowledge support from the WP 5.1 \textit{MHD\textsf{@}Exascale} (ANR-22-EXOR-0015) of the Programme et équipements prioritaires de recherche (PEPR) \textit{Origins} (PI: A. Morbidelli).  The SHAMROCK code follows the developments of the WP 5.1 of PEPR \textit{Origins}. 
We acknowledge funding from the ERC CoG project PODCAST No 864965. BC also acknowledges the funding from program ANR-20-CE49-0006 (ANR DISKBUILD) and funding from the European High Performance Computing Joint Undertaking (JU) under grant agreement No 101093441, supported by contributions from multiple member states, including ANR (ANR-23-EHPC-0007-02). We thank the referee whose comments, questions and suggestions help to improve the quality of the paper.\end{acknowledgements}

\bibliographystyle{aa} 
\bibliography{ref_source}

\begin{appendix} 
\section{Drag integrators}

To solve for Eqs.~(\ref{eq:CompEq}) -- (\ref{eq:CompVar4}), we first integrated  the transport and the source terms between the steps $\left( n \right)$ and $\left( n + \frac{1}{2}\right)$. We then solved for  
\begin{equation}
        \frac{\partial \Tilde{\bm{U}}}{\partial t} = \Vec{\Tilde{f}}_{\rm drag},
            \label{eq:DragEq_bis}
    \end{equation}
where
    \begin{equation}
        \Tilde{\bm{U}} \equiv \left[ \rho_{\mathrm{g}}\Vec{v}_{\mathrm{g}}, \rho_{\rm d,1}\Vec{v}_{\mathrm{d},1},..., \rho_{\mathrm{d},N}\Vec{v}_{{\mathrm{d}},N}\right]^{T} 
        = \left[\Vec{M}_{\mathrm{g}}, \Vec{M}_{\mathrm{d},1},..., \Vec{M}_{\mathrm{d},N} \right]^{T},
        \label{eq:DragEqVar1_bis}
    \end{equation}
and $\Vec{\Tilde{f}}_{\rm drag}$ is defined by Eq.~(\ref{eq:DragEq}), between steps $\left( n + \frac{1}{2}\right)$ and $\left( n + 1 \right)$. For IMEX(Implicit-Explicit)-like schemes, momenta terms should be modified accordingly for consistency. In the following, we review the different algorithms used to solve Eqs.~(\ref{eq:DragEq_bis}) -- (\ref{eq:DragEqVar1_bis}). 

\subsection{General first-order implicit RK1}
Denoting $\Delta \Tilde{U} =\Tilde{\bm{U}}^{(n+1)} - \Tilde{\bm{U}}^{(n)}$, and integrating Eq. ~(\ref{eq:DragEq_bis}) from $ t = t^{n}$ to $t = t^{n+1}$ provides the first-order implicit RK1 method and 

\begin{equation}
    \begin{split}
        \frac{\Delta \Tilde{U}  }{\Delta t} &= \Vec{\Tilde{f}}_{\rm drag}\left(\Tilde{\bm{U}}^{(n +1)} \right)\\
        &= \Vec{\Tilde{f}}_{\rm drag}\left(\Tilde{\bm{U}}^{(n)} + \left( \Tilde{\bm{U}}^{(n + 1)} - \Tilde{\bm{U}}^{(n)}\right) \right)\\
        &\approx \Vec{\Tilde{f}}_{\rm drag}\left(\Tilde{\bm{U}}^{n )} \right) + \Vec{\Omega} \Delta \Vec{\Tilde{U}}
    \end{split},
    \label{eq:eq_rk1_0}
\end{equation}
where the Jacobian matrix $\Vec{\Omega} $  is given by Eq.~(\ref{eq:jacobian}). Corrections for IMEX-like schemes are taken into account by replacing $\Vec{\Tilde{f}}_{\rm drag}$ in  Eq.~(\ref{eq:DragEq_bis}) by $\Vec{\Tilde{f}}$ given by
\begin{equation}
    \Vec{\Tilde{f}} =\Vec{\Tilde{f}}_{\rm drag} + \frac{\left( \Tilde{\bm{U}}^{(n+\frac{1}{2})} - \Tilde{\bm{U}}^{(n)}\right) }{\Delta t}
    \label{eq:eq_rk1_1} ,
\end{equation}
where $\Tilde{\bm{U}}^{(n)}$, $\Tilde{\bm{U}}^{(n+\frac{1}{2})}$ denote the momenta vectors before and after the hydrodynamic steps respectively. As such, one solves for
\begin{equation}
   \Tilde{\bm{U}}^{(n + 1)} - \Tilde{\bm{U}}^{(n)} = \left( \bm{I} - \Delta t \Vec{\Omega} \right)^{-1} \left[ \Delta t \Vec{\Tilde{f}}_{\rm drag}\left( \Tilde{\bm{U}}^{(n)}\right) +  \left( \Tilde{\bm{U}}^{(n+\frac{1}{2})} - \Tilde{\bm{U}}^{(n)}\right) \right]
    \label{eq:eq_rk1_2}.
\end{equation}

For $N$ dust fluids, Eq.~(\ref{eq:eq_rk1_2}) can be solved using direct methods with a time complexity of $\mathcal{O}\left( N^{3} \right)$ such as PLU decomposition. This algorithm was used by \cite{Benitez_Llambay_2019}, only for testing purposes.

\subsection{Implicit RK1 with analytical solution}
\label{app:analytical}

A more effective alternative of complexity $\mathcal{O}\left( N\right)$ consists of using an analytical solution of the sparse matrix Eq.~(\ref{eq:eq_rk1_2}). For a given vector $\bm{b} = \left(b_0, b_1, ..., b_N \right)$, we aim to find the vector $\bm{q} =  \left(q_0,q_1, ..., q_N \right)$ such that 
\begin{equation}
    \left( \bm{I} - \Delta t \Vec{\Omega} \right)\bm{q} = \bm{b}.
    \label{eq:analytic_1}
\end{equation}
With some algebra

\begin{equation}
\left( \ref{eq:analytic_1} \right) \iff 
\begin{cases}
\left( 1 + \sum_{i =1}^{N} \epsilon_i \alpha_i \Delta t \right) q_0 
- \sum_{i=1}^{N} \alpha_i \Delta t \, q_i = b_0, \\[6pt]

- \epsilon_1 \alpha_1 \Delta t \, q_0 
+ \left( 1 + \alpha_1 \Delta t \right) q_1 = b_1, \\[6pt]

\quad \vdots \\[6pt]

- \epsilon_N \alpha_N \Delta t \, q_0 
+ \left( 1 + \alpha_N \Delta t \right) q_N = b_N.
\end{cases}
\label{eq:analytic_2}
\end{equation}

From lines $2$--$\left(N +1\right)$ of Eq.~(\ref{eq:analytic_2}) we have for $i = 1,\dots,N$ 
    \begin{equation}
        q_i = \frac{1}{\left(1 + \alpha_i \Delta t  \right)} \left[ b_i + \left(\epsilon_i \alpha_i \Delta t \right) q_0\right].
        \label{eq:analytic_3}
    \end{equation}
    Replacing Eq.~(\ref{eq:analytic_3}) back into the first line of Eq.~(\ref{eq:analytic_2}) gives
    \begin{equation}
         q_0 = \frac{b_0 + \Delta t \mathcal{B}}{\mathcal{A}},
        \label{eq:analytic_4}
    \end{equation}
    with $\mathcal{A} = 1 + \sum_{i=1}^{N} \left( 1 - \frac{1}{1 +\alpha_i \Delta t  }\right) \epsilon_i$ and $\mathcal{B} = \sum_{i=1}^{N} \left( \frac{\alpha_i}{1 + \alpha_i \Delta t  }\right)b_i $.

\subsection{Second-order implicit RK2}
The second-order implicit RK2 drag solver proposed by \citet{Huang_2022} is consistent with the two-stage second-order Total Variation Diminishing (TVD) Runge-Kutta scheme of \citet{Gottlieb_Shu}. 
The first stage of the drag integration step consists of a first-order implicit RK1. The second stage of the drag integration step is described as follows
\begin{equation}
    \Tilde{\bm{U}}^{(n + 1)} - \Tilde{\bm{U}}^{(n)} = \frac{\Delta t}{2}\Vec{\Tilde{f}}\left(\Tilde{\bm{U}}^{(n)}, \Tilde{\bm{V}}^{(n)}\right) + \frac{\Delta t}{2}\Vec{\Tilde{f}}\left(\Tilde{\bm{U}}^{(n+1)}, \Tilde{\bm{V}}^{(n+\frac{1}{2})}\right) ,
    \label{eq:rk_2_0}
\end{equation}

where

\begin{equation}
\Tilde{\bm{V}} \equiv \left( \epsilon , \alpha \right),
\label{eq:rk_2_0_1}
\end{equation}
with $\epsilon = \left[ \epsilon_1, \cdots, \epsilon_n \right]$ and $\alpha = \left[ \alpha_1, \cdots, \alpha_n \right]$.

From a Taylor expansion in $\vert\vert \Vec{\Omega} \vert \vert \Delta t $, we have
\begin{equation}
    \begin{split}
        \Vec{\Tilde{f}}\left(\Tilde{\bm{U}}^{(n)}, \Tilde{\bm{V}}^{(n)}\right) &= \Vec{\Tilde{f}}\left(\Tilde{\bm{U}}^{(n+1)} - \Delta t \Vec{\Tilde{f}}\left(\Tilde{\bm{U}}^{(n+1)}, \Tilde{\bm{V}}^{(n)}\right), \Tilde{\bm{V}}^{(n)}\right) \\
         &=  \left(\bm{I}- \Delta t \Vec{\Omega}\left(\Tilde{\bm{U}}^{(n+1)}, \Tilde{\bm{V}}^{(n)}\right) \right) \Vec{\Tilde{f}}\left(\Tilde{\bm{U}}^{(n+1)}, \Tilde{\bm{V}}^{(n)}\right)\\
         &= \left(\bm{I}- \Delta t \Vec{\Omega}\left(\Tilde{\bm{U}}^{(n+1)}, \Tilde{\bm{V}}^{(n)}\right) \right) 
         \left(       \Vec{\Tilde{f}}\left(\Tilde{\bm{U}}^{(n)}, \Tilde{\bm{V}}^{(n)}\right)+
         \right. \\
         \left. \Vec{\Omega}\left(\Tilde{\bm{U}}^{(n)}, \Tilde{\bm{V}}^{(n)}\right) \Delta  \Tilde{\bm{U}} \right). \\
    \end{split}
    \label{eq:rk_2_1}
\end{equation}
 Similarly,
\begin{equation}
    \Vec{\Tilde{f}}\left(\Tilde{\bm{U}}^{(n+1)}, \Tilde{\bm{V}}^{(n+\frac{1}{2})}\right) = \Vec{\Tilde{f}}\left(\Tilde{\bm{U}}^{(n)}, \Tilde{\bm{V}}^{(n+\frac{1}{2})}\right)+ \Vec{\Omega}\left(\Tilde{\bm{U}}^{(n)}, \Tilde{\bm{V}}^{(n+\frac{1}{2})}\right) \Delta  \Tilde{\bm{U}} .
    \label{eq:rk_2_2}
\end{equation}

Replacing Eqs.~(\ref{eq:rk_2_1}) -- ~(\ref{eq:rk_2_2}) into Eq.~(\ref{eq:rk_2_0}) gives
\begin{equation}
\begin{split}
     \Delta  \Tilde{\bm{U}} = \frac{\Delta t}{2}\bm{\Lambda}^{-1}\left[\Vec{\Tilde{f}}\left(\Tilde{\bm{U}}^{(n)}, \Tilde{\bm{V}}^{(n+\frac{1}{2})}\right) + \left(\bm{I}- \Delta t \Vec{\Omega}\left(\Tilde{\bm{U}}^{(n+1)}, \Tilde{\bm{V}}^{(n)}\right) \right)  \right. \\
         \left. \Vec{\Tilde{f}}\left(\Tilde{\bm{U}}^{(n)}, \Tilde{\bm{V}}^{(n)}\right) \right],
\end{split}
 \label{eq:rk_2_3}
\end{equation}
where 
\begin{equation}
    \bm{\Lambda} \equiv \bm{I} - \frac{\Delta t}{2} \left(\bm{I} -  \Delta t \Vec{\Omega}\left(\Tilde{\bm{U}}^{(n+1)}, \Tilde{\bm{V}}^{(n)}\right) \right) \Vec{\Omega}\left(\Tilde{\bm{U}}^{(n)}, \Tilde{\bm{V}}^{(n)}\right) -\Vec{\Omega}\left(\Tilde{\bm{U}}^{(n)}, \Tilde{\bm{V}}^{(n+\frac{1}{2})}\right).
    \label{eq:rk_2_4}
\end{equation}
The inverse of $\bm{\Lambda}$ is obtained by a PLU decomposition.

\subsection{Second-order diagonally implicit Runge-Kutta MDIRK}

As an alternative to operator splitting, \citet{Krapp_2024} suggest combining the explicit and implicit terms in a single time step when integrating from a step $n$ to a step $n+1$, as 

    \begin{equation}
            \bm{U^{(*)}} = \bm{U}^{(n)} + \Tilde{\gamma} \Delta t \mathcal{L} \left(\bm{U}^{(n)} \right),
            \label{eq:mdirk1}
        \end{equation}
        
        \begin{equation}
            \bm{U^{(**)}} = \bm{U}^{(n)} + (1 - \delta) \Delta t \mathcal{L} \left(\bm{U}^{(*)} + \gamma \Delta t \bm{K}^{(*)} \right),
             \label{eq:mdirk2}
        \end{equation}
    
        \begin{equation}
            \bm{U^{(p)}} = \bm{U}^{(**)} + \frac{\delta}{\Tilde{\gamma}} \left(\bm{U}^{(*)} - \bm{U}^{(n)} \right) + b_2 \Delta t \mathbf{K}^{(p)},
             \label{eq:mdirk3}
        \end{equation}
        
        \begin{equation}
            \bm{U^{(n+1)}} = \bm{U}^{(p)} + \left(b_1 - \beta \right)\Delta t \bm{K}^{(*)} + b_2 \Delta t \bm{K}^{(p)} ,
             \label{eq:mdirk4}
        \end{equation}
where
\begin{equation}
\bm{U} \equiv
\end{equation}
is defined for both gas and dust. The operator       
        \begin{equation}
            \mathcal{L} (\bm{U}^{(n)}_{i}) = -\frac{1}{dx} \left( \bm{F}_{i+1/2}(\bm{U^{(n)}}) - \bm{F}_{i-1/2}(\bm{U^{(n)}})\right),
             \label{eq:mdirk5}
        \end{equation}
corresponds to the transport step. It returns flux divergence as described by Eq.~(\ref{eq:mdirk5}) and $\Tilde{\gamma} = \gamma$, $\delta = 1 - 1 / (2\gamma)$, $\beta = 1 - \gamma$, $b_1 = 1 - \gamma$, $b_2 = \gamma$. $\bm{K}^{(p)}$ and $\bm{K}^{(*)}$ are obtained by solving the system
\begin{align}
 \left(\bm{I} - \gamma \Delta t \Vec{\Omega} \right)\bm{K}^{(p)} & = \Vec{\Omega} M^{(p)},
             \label{eq:mdirk6} \\
 \left(\bm{I} - \gamma \Delta t \Vec{\Omega} \right)\bm{K}^{(*)} & = \Vec{\Omega} M^{(*)}.
             \label{eq:mdirk7}             
\end{align}
where 
\begin{equation}
    \bm{K} \equiv \left( k_{\mathrm{g}}^n, k_{\mathrm{d},1}^n, \cdots, k_{\mathrm{d},N-1}^n \right).
    \label{eq:mdirk8}
\end{equation}

The resolution of Eqs.~(\ref{eq:mdirk6}) -- (\ref{eq:mdirk8}) can be performed analytically, similarly to Appendix~\ref{app:analytical}.

\section{ Modified Taylor-based scaling and squaring algorithm}\label{app:algo_exp}

We explain the different steps involved in the modified Taylor-based scaling and squaring algorithm of \citet{Sastre2015} described in Sect.~\ref{sec:series}.

\subsection{Backward error analysis}

The choice of polynomial order $m$ and the scale factor $s$ is based on a backward error analysis, given by the following theorem \citep{Sastre2015}.

\begin{theorem}
    Let the matrix exponential Taylor approximation $ T_m \left(\bm{A} \right)$ of a given matrix $\bm{A}$ as defined by Eq.~(\ref{eq:ExpMatEqu}),
    satisfy 
        \begin{equation}
            \mathrm{e}^{-2^{-s} \bm{A}}T_m(2^{-s}\bm{A})= \bm{I} + \bm{G},
            \label{eq:eq_B1}
        \end{equation}
        with $ \left\lVert \bm{G} \right\rVert < 1$. Then 
        \begin{equation}
            \left[T_m(2^{-s}\bm{A})\right]^{2s} = \mathrm{e}^{\bm{A}+\bm{E}},
            \label{eq:eq_B2}
        \end{equation}
        where $\bm{E}$ commutes with $\bm{A}$ and 
        \begin{equation}
            \frac{ \left\lVert \bm{E} \right\rVert}{ \left\lVert \bm{A} \right\rVert} \leq \frac{-\log\left(1- \left\lVert \bm{G} \right\rVert \right)}{ \left\lVert 2^{-s}\bm{A} \right\rVert}.
            \label{eq:eq_B3}
        \end{equation}
\end{theorem}

\vspace{0.5cm}
The residue $\bm{E}$ quantifies the error between the exact exponential of the matrix and its numerical approximation. To improve the estimate, one should minimize the right-hand side of Eq.~(\ref{eq:eq_B3}). This term can be expressed as a function of $\theta \equiv \left\lVert 2^{-s}\bm{A}  \right\rVert$. For this purpose, define the function 
    \begin{equation}
        \rho(x) = \mathrm{e}^{-x}T_m(x) - 1.
        \label{eq:eq_B4}
    \end{equation}
One has
    \begin{equation}
        \begin{split}
            \rho(x) &= - \mathrm{e}^{-x}[\mathrm{e}^{x} - T_m(x)] \\
                    &=  - \mathrm{e}^{-x}\left( \sum_{j\geq m +1}\frac{x^{j}}{j!} \right) \\
                    &= \left( \sum_{j\geq 0}\frac{(-1)x^{j}}{j!} \right) \left( \sum_{j\geq m +1}\frac{x^{j}}{j!} \right)\\
                    &= \sum_{j \geq m+1} c_{j} x^{j}.
        \end{split}
    \end{equation}
From Eq.~(\ref{eq:eq_B1}),  
    \begin{equation}
        \left\lVert \bm{G} \right\rVert  = \left\lVert \rho(2^{-s} \bm{A}) \right\rVert \leq \sum_{j \geq m+1} |c_{j}| \theta^{j} \equiv f_m \left(\theta \right).
        \label{eq:eq_B6}
    \end{equation}
Combining Eqs.~(\ref{eq:eq_B3}) and (\ref{eq:eq_B3}) gives the bound 
    \begin{equation}
        \frac{ \left\lVert \bm{E} \right\rVert}{ \left\lVert \bm{A} \right\rVert} \leq \frac{-\log \left(1-f_m \left(\theta \right) \right)}{ \theta}.
        \label{eq:eq_B7}
    \end{equation}

 Using a zero-finder tool (symbolic Math Toolbox of MATLAB), we determined the highest value of $\theta \equiv \left\lVert 2^{-s}A  \right\rVert$ for a range of $m$ such that the bound is at most equal to $u = 2^{-53}$ (the unit round-off in IEEE double precision arithmetic).
    \begin{equation}
        \theta_{m} \equiv \max \{ \theta : \frac{-\log \left(1-f_m \left(\theta \right) \right)}{ \theta} \leq u \},
        \label{eq:eq_B8}
    \end{equation}
     $u = 2^{-53}$ being the unit round-off in IEEE double precision arithmetic. One can take another value depending on the desired accuracy : $2^{-11}$, $2^{-24}$, $2^{-113}$ and $2^{-237}$ for half, single, quadruple and octuple precision arithmetic respectively.

\subsection{Numerical cost}

    In the evaluation step, the matrix $T_{m}\left(2^{-s} \bm{A} \right)$ is calculated from Eq.~(\ref{eq:eq_B1}). The matrix polynomial is computed with the {Horner-Paterson-Stockmeyer} method to minimize the number of matrix products involved \citep{High05}.
    In the rest of the paper we used
    \begin{equation}
        \Pi_{m} \equiv \left \lfloor  \sqrt{m} \right  \rfloor +  \left \lfloor  m/\sqrt{m} \right  \rfloor,
        \label{eq:eval_squared_cost}
    \end{equation}
    the optimal number of matrix multiplications in the scaling phase with a Taylor approximation of order $m$, where $\lfloor x \rfloor$ denotes the integer part of the real number $x$. One defines the scaling factor $s$ by
    \begin{equation}
        s \equiv \begin{cases}
            \left \lfloor  \log_{2} (\left \lVert \bm{A} \right \rVert / \theta_{m} ) \right  \rfloor , & \left \lVert \bm{A} \right \rVert \geq \theta_{m} \\

            0 &, \left \lVert \bm{A} \right \rVert < \theta_{m}
         \end{cases}.
         \label{eq:scale_factor}
    \end{equation}
    
    The total cost in terms of matrix multiplications to compute the order-$m$ Taylor-based matrix exponential approximant is \citep{High05}
    \begin{equation}
    \Pi_{m} + s = \Pi_m + \max\left( \left \lfloor  \log_{2} \left(\left \lVert \bm{A} \right \rVert / \theta_{m} \right) \right  \rfloor , 0 \right) .
    \end{equation}
    The optimal truncate order of the Taylor polynomial is the one that minimizes the quantity $\Pi_m + s$. For $\left \lVert \bm{A} \right \rVert \geq \theta_{m}$, the optimal $m$ is such that $\Gamma_{m} = \Pi_{m} - \log_{2}(\theta_m)$ is the minimum cost, since $\left \lVert \bm{A} \right \rVert$ does not depend on $m$.
    
    \begin{table}
        \centering
        \begin{tabular}{cccccccc}
           \hline\hline
           $k$ & $m_k$ & $q_k$ & $r_k$ & 
           $\theta_{m_k}$ & $\theta_{m_k}^{'}$ & $\Pi_{m}$\\
         \hline  1& $2$& $1$  & $2$  & $2.5810 \times 10^{-8}$&  $8.7334 \times 10^{-6}$ & $1 $ \\
             2& $4$& $2$ & $2$ & $3.3972 \times 10^{-4}$  & $1.6778 \times 10^{-3}$ &  $2$ \\
             3& $6$& $2$ & $3$ & $9.0657 \times 10^{-3}$ &  $1.7720 \times 10^{-2}$&  $3$ \\
             4& $9$& $3$ & $3$ & $8.9578 \times 10^{-2}$ &  $1.1354 \times 10^{-1}$&  $4$ \\
             5& $12$& $3$ & $4$  & $2/9962\times 10^{-1}$ &  $3.2690\times 10^{-1}$&  $5$ \\
             6& $16$&  $4$& $4$ & $7.80\times 10^{-1}$ &  $7.8738\times 10^{-1}$& $6$  \\
             7& $20$& $4$ & $5$ &$1.4383 $  & $1.4383$ &  $7$ \\
             8& $25$& $5$ &$5$  & $2.4286$ & $2.4286$ &$8$  & \\
             9& $30$& $5$ & $6$  &  $3.5397$& $3.5397$ & $9$  \\
        \hline
        \end{tabular}
        \caption{Precomputed parameters for Taylor-based scaling and squaring algorithm for $ u = 2^{-53}$. In each row from left to right we have respectively the number of matrix product in the scaling phase $k$, the Taylor series polynomial order $m_k$, the Horner-Paterson-Stockmeyer parameters $q_k$ and $r_k$, and the number of matrix product $\Pi_{m_k}$. }
        \label{tab:params_table}
    \end{table}

    \subsection{First modification for choosing $m$ and $s$}
    
    For a maximum number $K$ of matrix product in the scaling step, we are looking for $k^*$ (optimal number of matrix product), $m^*$ (optimal Taylor polynomial order associated with $k^*$) and $s^*$ (optimal scale factor). The total cost is then $k^* + s^*$. 
 
    The algorithm ~\ref{algo_order_scale} presents the different strategy to choose the optimal order $m^*$ and the optimal scale factor $s^*$. The algorithm~\ref{algo_order_scale} with option 0 is the standard algorithm to choose $m^*$ and $s^*$. The first modification proposed by \citeauthor{Sastre2015} is to reduce the total cost $\left(k^* + s^* \right)$ in some cases (option 1 and option 2 in algorithm ~\ref{algo_order_scale}). Option 1 involves decreasing the number of matrix products by one, while option 2 involves the cost of additional squaring to select $k^* < K$ so that the overall cost is less than one given by option 1.

    The high number of branching logic in this algorithm is subject to a high rate of \textit{branch divergence} on GPUs. So, we rewrote algorithm ~\ref{algo_order_scale} for option 2 to reduce the number of if-else statements, and then reduce \textit{branch divergence}. The pseudo-code for the new algorithm is presented in algorithm. \ref{algo_order_scale_new}.
  
    \begin{algorithm}
    \caption{Polynomial order and scale factor choice } \label{algo_order_scale}
    \SetAlgoLined
    \SetKwInOut{Input}{input}
    \SetKwInOut{Output}{output}
    \DontPrintSemicolon
    \Input{
       option $:=0,1$, or $2$;\\
       $K$ maximum number of matrix multiplications allowed in scaling phase;\\
       $\{ m_k \}$ and $\{\theta_{m_k} \}$ from table ~\ref{tab:params_table} \\
    }

   \Output{$k^{*}$, $m^{*}$, and $s^*$.}

    \BlankLine
    $m^* = m_K$\;
    $k^* = K$\;
      $\hat{s} =  \left \lceil  \log_{2} (\left \lVert \bm{A} \right \rVert / \theta_{m_K} ) \right  \rceil$ \
    \eIf{($\left \lVert \bm{A} \right \rVert  \leq \theta_{m_K}$)}{
    $s^* = 0$; $\hat{s} = 0$\;
    \For{($k = 1 : K$)}
    {
        \If{($\left \lVert \bm{A} \right \rVert \leq \theta_{m_k}$ )}
        {
        $m^* = m_k$\;
        $k^* = k$\;
        break \;
        }
    }

    \If{option $=2$}
    {
        \uIf{($k \geq 8 $ and $\left \lVert \bm{A} \right \rVert \leq 2 \theta_{m_{k-2}}$) }
        {
            $k^* = k -2 $\;
            $m^* = m_{k^*}$\;
            $s^* = 1$;
        }\uElseIf{($k =9$ and $\left \lVert \bm{A} \right \rVert \leq 4\theta_{m_{k-3}}$)}
        {
            $k^* = k -3 $\;
            $m^* = m_k $\;
            $s^* = 2 $\;
        }
        
    }
    
    }
    {
        $\hat{s} =  \left \lfloor  \log_{2} (\left \lVert \bm{A} \right \rVert / \theta_{m_K} ) \right  \rfloor$ \;
        $s^* = \hat{s}$ \;
        \If{ option $>0$}
        {
            \uIf{$K \geq 7$ and $\frac{\left \lVert \bm{A} \right \rVert}{2^{\hat{s}}} \leq \theta_{m_{K-1}}$}
            {
                $k^* = K - 1$ \;
                $m^* = m_{k^*}$\;
            }
            \uElseIf{option $=2$}
            {
                \uIf{$K \geq 8$ and  $\frac{\left \lVert \bm{A} \right \rVert}{2^{\hat{s}}} \leq 2\theta_{m_{K-2}}$}
                {
                    $k^* = K -2$\;
                    $m^* =m_{k^*}$ \;
                    $s^* = \hat{s} + 1$\;
                }
                \uElseIf{ $ K =9$ and $ \frac{\left \lVert \bm{A} \right \rVert}{2^{\hat{s}}} \leq 4\theta_{m_{K-3}} $}
                {
                    $k^* = K -3$\;
                    $m^* =m_{k^*}$ \;
                    $s^* = \hat{s} + 2$\;
                }
            }
        }
    }
    \end{algorithm}

    \begin{algorithm}
    \caption{Modified polynomial order and scale factor choice} \label{algo_order_scale_new}
    \SetAlgoLined
    \SetKwInOut{Input}{input}
    \SetKwInOut{Output}{output}
    \DontPrintSemicolon
    \Input{
       $K$ maximum number of matrix multiplications allowed in scaling phase;\\
       $\{ m_k \}$ and $\{\theta_{m_k} \}$ from table ~\ref{tab:params_table} \\
    }

   \Output{$k^{*}$, $m^{*}$, and $s^*$.}

    \BlankLine
    $m^* = m_K$\;
    $k^* = K$\;
    $\hat{s} = \left \lceil \max \left (  \log_{2} (\left \lVert \bm{A} \right \rVert / \theta_{m_K} ) , 0 \right) \right \rceil$\;
    $s^* = 0$\;
    $cond = FALSE$\;
    \For{($k = 1 : K$) and NOT $cond$}
    {
        $cond = \left \lVert \bm{A} \right \rVert \leq \min \left(\theta_{m_k},\theta_{m_K}\right) $\;
        $m^* = cond * m_k + \left(1-cond \right) * m^*$  \;
        $k^* = cond * k + \left(1-cond \right) * k^*$\;
    }

    $\hat{k} = \min\left( k,K\right)$\;
    $cmp = \frac{\left \lVert \bm{A} \right \rVert}{2^{\hat{s}}}$\;
    $v_1 = \left[ \left( \hat{k} \geq 7 \right) and \left( cmp \leq \theta_{m_{\hat{k} - 1}} \right)\right]$\;
    $v_2 = \left[ \left( \hat{k} \geq 8 \right) and \left( cmp \leq 2 * \theta_{m_{\hat{k} - 2}} \right)\right] * 2$\;
    $v_3 = \left[ \left( \hat{k} \geq 9 \right) and \left( cmp \leq 4*\theta_{m_{\hat{k} - 3}} \right)\right] * 3 $\;
    $v = \max\left( v1,v2,v3\right)$\;

    \uIf{$v = v_1$}
    {
        $k^* = \hat{k} -1$; $m^* = m_k^*$\;
    }
    \uElseIf{$v = v_2 $}
    {
         $k^* = \hat{k} -2$; $m^* = m_k^*$; $s^* = \max\left( 1,\hat{s} + 1\right)$\;
    }
    \uElseIf{$v = v_3$}
    {
        $k^* = \hat{k} - 3$; $m^* = m_k^*$; $s^* = \max\left( 2,\hat{s} + 2\right)$\;
    }

    \end{algorithm}

    \subsection{Second modification}

    To evaluate the matrix polynomial of $m$ order  in Eq.~(\ref{eq:ExpMatEqu}), the Paterson-Stockmeyer method \citep{Pat_Stk} consists of choosing an integer $q < m$ and writing the matrix polynomial as a polynomial of degree $r = \left \lfloor \frac{m}{q} \right  \rfloor$ in $\bm{A}^q$, such that

    \begin{equation}
        T_m\left(\bm{A}\right) = \sum_{i=0}^{r}\bm{B}_i (\bm{A}^q)^i, 
        \label{eq:Pat_Stk}
    \end{equation}
    where $\bm{B}_i$, $i=0$,...,$r$ are matrix polynomials of degrees $q$ 
    \begin{equation}
        \bm{B}_i \equiv \sum_{j=0}^{q-1} b_{qi + j}\bm{A}^{j},
        \label{eq:Pat_stk2}
    \end{equation}
    and 
    \begin{equation}
        \bm{B}_r \equiv b_{rq} \bm{I} + ... + b_m \bm{A}_{m-qr}.
        \label{eq:Pat_Stk3}
    \end{equation}

    The algorithm \ref{alg_ps_coef} presents the procedure for evaluating the Paterson-Stockmeyer coefficients $r$ $\{ \bm{B}_{k} \}$.

     \begin{algorithm}
        \caption{Paterson-Stockmeyer coefficients} \label{alg_ps_coef}
        \SetAlgoLined
        \SetKwInOut{Input}{input}
        \SetKwInOut{Output}{output}
        \DontPrintSemicolon
        \Input{
           $q$; $r$;\;
           $\{\bm{A}_j = \bm{A}^j\}$, $j = 1:q$;\;
           ${b_i = (1)^i/i!}, i=0 :qr$;\;
           
        }
        \Output{$\{ \bm{B}_i\}, i = 0 : r-1$}

        \BlankLine
        \For{ $k= 0 : r-1 $}
        {
            $\bm{B}_k = 0$ \;
            \For{ $ j = 1 : q$}
            {
                $\bm{B}_k = \bm{B}_k + b_{qk + j} \bm{A}_j$ \;
            }
        }        
    \end{algorithm}

Evaluating the matrix polynomial from Eq.~(\ref{eq:Pat_Stk}) using the Paterson-Stockmeyer algorithm in combination with the Horner technique reduces the number of matrix products \citep{High05}. It is called the Horner-Paterson-Stockmeyer method. Equation~(\ref{eq:Pat_Stk3}) shows that when $q$ divides $m$, one has $\bm{B}_r = b_m \bm{I}$. The values of $q$ are given in the third column of table~\ref{tab:params_table}. And $m_{k+1} = m_k + q_k$. Thus, a reduction of the number of matrix products by 1 reduces the order of the Taylor polynomial by $q_k$. Based on this remark, when the inequality

    \begin{equation}
        \frac{\left \lVert T_{m_k}\left( \bm{A}\right) -  T_{m_{k-1}}\left( \bm{A} \right)  \right \rVert}{\left \lVert \mathrm{e}^{\bm{A}} \right \rVert } \leq u,
        \label{eq:Pat_Stk4}
    \end{equation}
   holds,  \citeauthor{Sastre2015} proposed to neglect the $q$ highest terms in $T_{m_k}\left(\bm{A} \right)$. This condition is verified by performing a numerical test, which consists of checking if the inequality 
    \begin{equation}
        \left \lVert \mathrm{e}^{\bm{-A}} \right \rVert \left \lVert \Bar{B}_{r-1} \right \rVert \left \lVert \bm{A}^{q} \right \rVert ^{r-1} \leq u ,
        \label{eq:Pat_Stk5}
    \end{equation}
    is satisfied. To do so, one estimates a bound for $ \left \lVert \mathrm{e}^{\bm{-A}} \right \rVert$ according to
    \begin{equation}
         \left \lVert \mathrm{e}^{\bm{-A}} \right \rVert \approx  \left \lVert T_{m_{k}} \left(\bm{-A} \right) \right \rVert \leq b_{\rm exp} =  \left \lVert b_0 \bm{I} + \hat{\bm{B}}_0 \right \rVert + \sum_{l=1}^{r-1}  \left \lVert \hat{\bm{B}}_l \right \rVert  \left \lVert \bm{A}^q \right \rVert ^{l},
         \label{eq:Pat_Stk6}
    \end{equation}
    where $\hat{\bm{B}}_l = \sum_{j=1}^{q} (-1)^{ql+j}b_{ql +j} \bm{A}^{j}$, $l$ = $0,...,r-1$.
    The condition given by Eq.~(\ref{eq:Pat_Stk5}) can therefore be replaced in practice by 
    \begin{equation}
         \left \lVert T_{m_k}\left( \bm{A}\right) -  T_{m_{k-1}}\left( \bm{A} \right)  \right \rVert \leq \left \lVert \Bar{B}_{r-1} \right \rVert \left \lVert \bm{A}^{q} \right \rVert ^{r-1},
         \label{eq:Pat_stk7}
    \end{equation}
    from Eq.~(\ref{eq:Pat_Stk5}). The numerical procedure to get $b_{\rm exp}$ the bound of $ \left \lVert \mathrm{e}^{\bm{-A}} \right \rVert$ is detailed in algorithm \ref{alg_bound}.
    
    \begin{algorithm}
        \caption{Compute bound $b_{exp}$} \label{alg_bound}
        \SetAlgoLined
        \SetKwInOut{Input}{input}
        \SetKwInOut{Output}{output}
        \DontPrintSemicolon
        \Input{
           $q$; $r$;\;
           $\{\bm{A}_j = \bm{A}^j\}$, $j = 1:q$;\;
           ${b_i = (-1)^i/i!}, i=0 :qr.$\;
        }
        \Output{$b_{exp}$}

        \BlankLine
        compute $\{ \hat{\bm{B}}_l\}, l =0: r-1$  $\longrightarrow $ Algo 2 \;
        $b_{exp} = \left \lVert \hat{\bm{B}}_{r-1} \right \rVert $\;
        \For{ $l =r-1 : -1: 2$}
        {
           $ b_{exp} = \left \lVert \hat{\bm{B}}_{r-1} \right \rVert + \left \lVert \bm{A}_{q} \right \rVert \times b_{exp} $\;
        }
        $b_{exp} = \left  \lVert b_0 \bm{I} + \hat{\bm{B}}_{0} \right \rVert + \left \lVert \bm{A}_{q} \right \rVert \times b_{exp} $\;
    \end{algorithm}

    Algorithm ~\ref{alg_hps_eval_2} presents the Horner-Paterson-Stockmeyer procedure to evaluate the Taylor polynomial or order $m$ during the scaling phase with the bound performance test to reduce the number of matrix products.

      \begin{algorithm}
        \caption{Taylor polynomial evaluation} \label{alg_hps_eval_2}
        \SetAlgoLined
        \SetKwInOut{Input}{input}
        \SetKwInOut{Output}{output}
        \DontPrintSemicolon
        \Input{
           $q$; $r$;\;
           $\{\bm{A}_j = \bm{A}^j\}$, $j = 1:q$;\;
           ${b_i = (1)^i/i!}, i=0 :qr$;\;
           $\{ \Bar{\bm{B}}_l\}, l =0: r-1$;\;
           $u$.
        }
        \Output{$F = T_m\left(\bm{A} \right)$}

        \BlankLine
        $F = \Bar{\bm{B}}_{r-1}$\;
        \For{ $j = r-1 : -1 : 1$}
        {
            \eIf{($b_{exp} \left \lVert \bm{F} \right \rVert \left \lVert \bm{A}_{q} \right \rVert ^ {j} \leq u$)}
            {
                $\bm{F} = \Bar{\bm{B}}_{j-1}$ \;
            }
            {
                $\bm{F} = \Bar{\bm{B}}_{j-1} + \bm{A}_{q} \times \bm{F}$  \;
            }
        }
        $\bm{F} = \bm{F} + b_0 \bm{I}$
        
    \end{algorithm}
Together, the two modifications are resumed in the algorithm ~\ref{alg_Tss}.

As mentioned in Sect.~\ref{scal_squ_sub_sec}, algorithm ~\ref{alg_new_poly} shows the new algorithm we proposed, which combines algorithms ~\ref{alg_ps_coef} and ~\ref{alg_hps_eval_2} (without bound tests) together.

 \begin{algorithm}
        \caption{Taylor-based scaling and squaring } \label{alg_Tss}
        \SetAlgoLined
        \SetKwInOut{Input}{input}
        \SetKwInOut{Output}{output}
        \DontPrintSemicolon
        \Input{
           $\bm{A}$, $K$, $u$\;
        }
        \Output{An approximation of $\mathrm{e}^{\left(\bm{A} \right)}$}

        \BlankLine
        Select optimal order of Taylor polynomial $m^*$ and optimal scaling factor  $s^*$ . $\longrightarrow $ Algorithm ~\ref{algo_order_scale} \;

        Perform Taylor polynomial evaluation for the scaled matrix $\bm{A}/2^s$ . $\longrightarrow $ Algorithm ~\ref{alg_hps_eval_2} \;

        Perform the appropriate number of squaring steps . \;

    \end{algorithm}

 \begin{algorithm}
        \caption{New Taylor polynomial evaluation without bound-test } \label{alg_new_poly}
        \SetAlgoLined
        \SetKwInOut{Input}{input}
        \SetKwInOut{Output}{output}
        \DontPrintSemicolon
        \Input{
           $\bm{A}$, $q$, $r$, $\{ b_i = \frac{1}{i!}\}, i=0:m$ \;
        }
        \Output{$T_m \left( \bm{A} \right) = b_0 \bm{I} + \sum_{k=0}^{r-1} \Bar{\bm{B}}_k \left( \bm{A}^q\right)^k$ .}

        \BlankLine
         
          $\bm{F} = \bm{0}$\;
          \For{$k = r-1 : -1 : 0 $}
          {
                $\Bar{\bm{B}}_k = \bm{0}$\;
                $I := $ identity matrix\;
                \For{$j = 1 : q$}
                {
                   $\bm{I} = \bm{A} $ x $\bm{I}$ $\longrightarrow $ $\left ( \bm{I} := \bm{A}^j\right)$\;
                   $\Bar{\bm{B}}_k = \Bar{\bm{B}}_k + b_{q*k + j} \bm{I}$ $\longrightarrow $ $\left ( \bm{I} := \bm{A}^j\right)$ \;
                }
                $cond = \left( k \geq 1\right)$\;
                $\bm{F} = \left( 1-cond \right)*  \left( \Bar{\bm{B}}_k + \bm{F}\right) + cond  $ * $\left( \Bar{\bm{B}}_k + \bm{F}\right)$ x $\bm{I}$  $\longrightarrow $ $\left ( \bm{I} := \bm{A}^q\right)$\;
          }
          $\bm{F} = \bm{F} +  b_0 \bm{I}$  $\longrightarrow $ $I := $ identity matrix\;
        
    \end{algorithm}

     \begin{algorithm}
        \caption{New compute bound $b_{exp}$} \label{alg_bound}
        \SetAlgoLined
        \SetKwInOut{Input}{input}
        \SetKwInOut{Output}{output}
        \DontPrintSemicolon
        \Input{
           $q$; $r$;\;
           $\bm{A}$ ;
           ${b_i = (-1)^i/i!}, i=0 :qr.$\;
        }
        \Output{$b_{exp}$}

        \BlankLine
        $b_{exp} = 0 $\;
        $\hat{\bm{B}} = \bm{0}$\;
        \For{ $l =r-1 : -1: 0$}
        {
            $\hat{\bm{B}} = \bm{0}$\;
            $I := $ identity matrix\;
            \For{$j = 1 : q$}
            {
               $\bm{I} = \bm{A} $ x $\bm{I}$ $\longrightarrow $ $\left ( \bm{I} := \bm{A}^j\right)$\;
               $\hat{\bm{B}} = \hat{\bm{B}} + b_{q*k + j} \bm{I}$ $\longrightarrow $ $\left ( \bm{I} := \bm{A}^j\right)$ \;
            }
            $cond = \left( k \geq 1 \right)$\;
           $ b_{exp} = cond *  \left( \left \lVert \hat{\bm{B}} \right  \rVert  + b_{exp} \right)  \times  \left \lVert \bm{I} \right \rVert + \left( 1- cond \right) * b_{exp}$\;
        }
        $b_{exp} = b_{exp} +  \left  \lVert b_0 \bm{I} + \hat{\bm{B}} \right \rVert $   $\longrightarrow $ $\left ( \hat{\bm{B}} := \hat{\bm{B}_0} \right)$ \;
    \end{algorithm}

 \begin{algorithm}
        \caption{New Taylor polynomial evaluation } \label{alg_new_poly}
        \SetAlgoLined
        \SetKwInOut{Input}{input}
        \SetKwInOut{Output}{output}
        \DontPrintSemicolon
        \Input{
           $\bm{A}$, $q$, $r$, $u$ $\{ b_i = \frac{1}{i!}\}, i=0:m$ \;
        }
        \Output{$T_m \left( \bm{A} \right) = b_0 \bm{I} + \sum_{k=0}^{r-1} \Bar{\bm{B}}_k \left( \bm{A}^q\right)^k$ .}

        \BlankLine
         
          $\bm{F} = \bm{0}$\;
          \For{$k = r-1 : -1 : 0 $}
          {
                $\Bar{\bm{B}}_k = \bm{0}$\;
                $I := $ identity matrix\;
                \For{$j = 1 : q$}
                {
                   $\bm{I} = \bm{A} $ x $\bm{I}$ $\longrightarrow $ $\left ( \bm{I} := \bm{A}^j\right)$\;
                   $\Bar{\bm{B}}_k = \Bar{\bm{B}}_k + b_{q*k + j} \bm{I}$ $\longrightarrow $ $\left ( \bm{I} := \bm{A}^j\right)$ \;
                }
                $cond_1 = \left( k \geq 1\right)$\;
                $cond_2 = \left( b_{exp} * \left  \lVert \bar{\bm{B}}  \right \rVert * \left  \lVert \bm{I}  \right \rVert ^{k} \right) \leq u $
                $\bm{F} = \left( 1-cond_1 \right)*  \left( \Bar{\bm{B}}_k + \bm{F}\right) + cond_1 \left(1-cond_2 \right)  $ * $\left( \Bar{\bm{B}}_k + \bm{F}\right)$ x $\bm{I}$  \;
          }
          $\bm{F} = \bm{F} +  b_0 \bm{I}$  $\longrightarrow $ $I := $ identity matrix\;
        
    \end{algorithm}
    
\end{appendix}
\end{document}